\documentstyle[12pt]{article}
\def\PR  #1 #2 #3 {Phys.~Rev.~{\bf #1}, #2 (#3)}
\def\PRL #1 #2 #3 {Phys.~Rev.~Lett.~{\bf #1}, #2 (#3)}
\def\PRD #1 #2 #3 {Phys.~Rev.~D~{\bf #1}, #2 (#3)}
\def\PLB #1 #2 #3 {Phys.~Lett.~{\bf B#1}, #2 (#3)}
\def\NPB #1 #2 #3 {Nucl.~Phys.~{\bf B#1}, #2 (#3)}
\def\RMP #1 #2 #3 {Rev.~Mod.~Phys.~{\bf #1}, #2 (#3)}

\begin{document}
\include{psfig}
\begin{titlepage}
 
\rightline{hep-ph/9604223}
\medskip
\rightline{June 1996}
\bigskip\bigskip

\begin{center}
{\Large\bf QCD and Yukawa corrections to single-top-quark production
via $q \bar q \to t \bar b$}\\
\bigskip\bigskip\bigskip\bigskip
{\large{\bf Martin C. Smith} and \bf{Scott S. Willenbrock}}\\
\medskip Department of Physics\\
University of Illinois\\
1110 West Green Street\\
Urbana, IL  61801\\ 
\end{center} 
\bigskip\bigskip\bigskip\bigskip

\begin{abstract}
We calculate the ${\sl O}(\alpha_s)$ and ${\sl O}(\alpha_W
m_t^2/M_W^2)$ corrections to the production of a single top quark via
the weak process $q \bar q \to t \bar b$ at the Fermilab Tevatron and
the CERN Large Hadron Collider.  An accurate calculation of the cross
section is necessary in order to extract $|V_{tb}|$ from experiment.
\end{abstract}

\addtolength{\baselineskip}{9pt}

\end{titlepage}

\section{Introduction}

\indent\indent The recent discovery of the top quark \cite{TOP} has
focused attention on top-quark physics.  With the advent of
accelerators able to produce copious numbers of top quarks, a
comparison of the top quark's observed properties with those predicted
by the Standard Model promises to be an important test of the model
and may well provide insight into exciting new physics.

\begin{figure}
\centerline{\psfig{figure=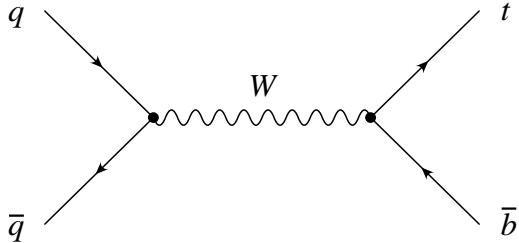,width=3in}}
\caption{\footnotesize Single-top-quark production via $q \bar q \to t
\bar b$.}
\label{tree graph}
\end{figure}

In this paper we calculate the next-to-leading-order cross section for
the weak process $q \bar{q} \to t \bar{b}$, which produces a single
top quark via a virtual $s$-channel $W$ boson (Fig.~\ref{tree graph})
\cite{CPet,SW}.  The most important corrections to the ${\sl
O}(\alpha_W^2)$ leading-order cross section are the QCD correction of
${\sl O}(\alpha_s)$ and the Yukawa correction of ${\sl O}(\alpha_W
m_t^2/M_W^2)$.  The Yukawa correction, which arises from loops of
Higgs bosons and the scalar components of virtual vector bosons,
dominates the ordinary ${\sl O}(\alpha_W)$ electroweak correction in
the large $m_t$ limit.  For the known value of the top-quark mass,
$m_t=175 \pm 6$ GeV, the Yukawa correction is expected to be at least
as large as the ordinary electroweak correction.

A precise theoretical calculation of the cross section for $q \bar q
\to t \bar b$ is necessary for a number of reasons. The cross section
obviously determines the yield of single top quarks produced via this
process.  More importantly, the coupling of the top quark to the $W$
boson in $q \bar q \to t \bar b$ is proportional to the
Cabibbo-Kobayashi-Maskawa (CKM) matrix element $V_{tb}$, one of the
few Standard Model parameters not yet measured experimentally.  If
there are only three generations, unitarity of the CKM matrix implies
that $|V_{tb}|$ must be very close to unity ($.9988 < |V_{tb}| <
.9995$) \cite{PDB}.  However, if there is a fourth generation,
$|V_{tb}|$ could be anything between (almost) zero and unity,
depending on the amount of mixing between the third and fourth
generations.  Measurement of the $q \bar q \to t \bar b$ cross
section, coupled with an accurate theoretical calculation, may provide
the best direct measurement of $|V_{tb}|$ \cite{SW}.  Finally, in
addition to being interesting in its own right, $q \bar q \to t \bar
b$ is a significant background to other processes, such as $q \bar q
\to W\!H$ with $H\to b\bar b$, where $H$ is the Higgs boson
\cite{HIGGS}.

\begin{figure}
\centerline{\psfig{figure=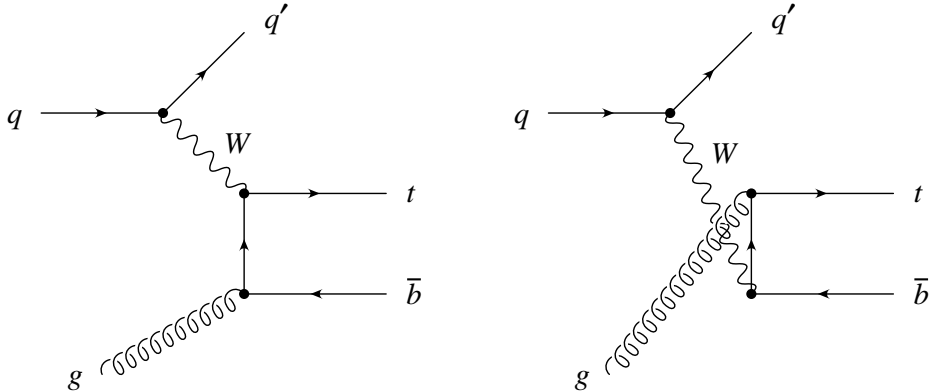,width=5in}}
\caption{\footnotesize Single-top-quark production via $W$-gluon
fusion.}
\label{Wg graph}
\end{figure}

In some ways, $q \bar q \to t \bar b$ is similar to the more-studied
$W$-gluon fusion process (Fig.~\ref{Wg graph}) \cite{WG}.  However,
where that process involves a space-like $W$ boson with $q^2 < 0$, the
process $q \bar q \to t \bar b$ proceeds via a time-like $W$ boson
with $q^2 > (m_t + m_b)^2$.  Thus these two processes, together with
the decay of the top quark, $t \to Wb$ (where the $W$ boson has $q^2
\approx M_W^2$), probe complementary aspects of the top quark's weak
charged current.  The kinematic distributions of the final-state
particles in the two processes also differ significantly. There is an
additional jet present in $W$-gluon fusion, and the $\bar b$ quark is
usually produced at low transverse momentum, while in $q \bar q \to t
\bar b$, the $\bar b$ quark recoils against the $t$ quark with high
transverse momentum.

At the Fermilab Tevatron ($\sqrt{S}$ = 2 TeV $p \bar p$ collider), the
sum of the cross sections for $q \bar q \to t \bar b$ and $q \bar q
\to \bar t b$ is roughly a factor of seven smaller than the dominant
$t \bar t$ production cross section \cite{TT}, and about a factor of
two smaller than the $W$-gluon-fusion cross section
\cite{WG}. Nevertheless, a recent study indicates that with double $b$
tagging, a signal is observable at the Tevatron with 2-3~fb$^{-1}$ of
integrated luminosity \cite{SW}.  Unfortunately, even though the $q
\bar q \to t \bar b, \bar t b$ cross section is larger at the CERN
Large Hadron Collider (LHC, $\sqrt{S}$ = 14 TeV $pp$ collider), the
signal will likely be obscured by backgrounds from the even larger $t
\bar t$ and $W$-gluon fusion processes, which are initiated by gluons
\cite{SW}.

An important feature of $q \bar q \to t \bar b$ is the accuracy with
which the cross section can be calculated.  The top-quark mass is much
larger than $\Lambda_{QCD}$, so calculations are performed in a regime
where perturbative QCD is very reliable.  The correction to the
initial state is identical to that occurring in the ordinary Drell-Yan
process $q \bar q \to W^*\!\to \bar \ell \nu$ ($W^*$ denotes a virtual
$W$ boson), which has been calculated to ${\sl O}(\alpha_s^2)$
\cite{DY2}.  Furthermore, by experimentally measuring $q \bar q \to
W^* \to \bar \ell \nu$, the initial quark-antiquark flux can be
constrained without recourse to perturbation theory.\footnote{Since
the longitudinal momentum of the neutrino cannot be reconstructed, the
$q^2$ of the $W^*$ cannot be determined, so $q \bar q \to W^* \to \bar
\ell \nu$ yields only a constraint on the quark-antiquark flux, rather
than a direct measurement.}  This provides a check of the parton
distribution functions, and allows the reduction of systematic errors.
The parton distribution functions are not expected to be a large
source of uncertainty, as the dominant contribution to the cross
section comes from quark and antiquark distribution functions
evaluated at relatively high values of $x$, where they are well known.
There is little sensitivity to the less-well-known gluon distribution
function, in contrast to the case of $W$-gluon fusion.  The
final-state correction to the inclusive cross section is
straightforward, and involves no collinear or infrared singularities.
The QCD corrections to the initial and final states do not interfere
at next-to-leading-order because the $t \bar b$ is in a color singlet
if a gluon is emitted from the initial state, but a color octet if it
is emitted from the final state.  There is, however, interference at
${\sl O}(\alpha_s^2)$ from the emission of two gluons.

This paper is organized as follows.  In Section 2 we present the
${\sl O}(\alpha_s)$ QCD corrections to both the initial and final
states, and discuss their dependence on the renormalization and
factorization scales.  In Section 3 we present the ${\sl O}(\alpha_W
m_t^2/M_W^2)$ Yukawa correction.  In Section 4 we present a summary of
our results.  We give an analytic expression for the Yukawa correction
in an appendix.

\section{QCD correction}

\begin{figure}
\centerline{\psfig{figure=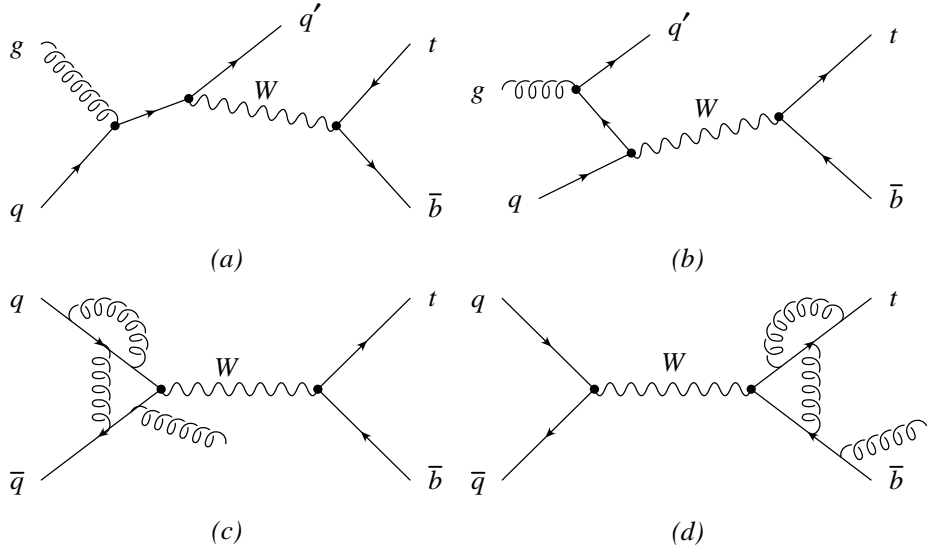,width=5in}}
\caption{\footnotesize ${\sl O}(\alpha_s)$ correction to $q \bar q
\to t \bar b$: (a)-(c) initial state, (d) final state.}
\label{QCD graphs}
\end{figure}

\indent\indent The diagrams which contribute to the ${\sl
O}(\alpha_s)$ correction to $q \bar q \to t \bar b$ are shown in
Fig.~\ref{QCD graphs}.  As mentioned in the Introduction, the QCD
corrections to the initial and final states do not interfere at ${\sl
O}(\alpha_s)$.  Therefore, we may consider the corrections to the
initial and final states separately.  To this end, we break up the
process $p \bar p \to t \bar b + X$ into the production of a virtual
$W$ boson of mass-squared $q^2$, followed by its propagation and decay
into $t \bar b$.

The production cross section of the virtual $W$ boson is formally
identical to that of the Drell-Yan process, to all orders in QCD. The
modulus squared of the decay amplitude of the virtual $W$ boson,
integrated over the phase space of all final-state particles, is
obtained by the application of Cutkosky's rules \cite{CUT} as twice
the imaginary part of the self-energy of the $W$ boson due to a $t
\bar b$ loop, again to all orders in QCD.  Furthermore, because the
current to which the $W$ boson couples in the initial state is
conserved to all orders in QCD (for massless quarks), we need only
consider the $-g^{\mu\nu}$ term in the $W$-boson propagator and
self-energy.  Thus we may write the differential cross section as
\begin{equation}
\frac{d\sigma}{dq^2}(p\bar{p} \to t\bar{b}+X) = \sigma(p\bar{p}\to
W^*\!+X) \frac{{\rm Im}\:\Pi(q^2,m_t^2,m_b^2)}{\pi(q^2-M_W^2)^2}
\label{MASTER}
\end{equation} 
where $\Pi$ is the coefficient of the $-g^{\mu\nu}$ term of the
self-energy of a $W$ boson with mass-squared $q^2$.  The total cross
section is obtained by integrating over $q^2$.  This equation is valid
to ${\sl O}(\alpha_s)$, but not beyond, because it neglects the
interference between the QCD corrections to the initial and final
states.

To demonstrate this procedure, we obtain the leading-order cross
section for $p \bar p \to t \bar b$ using
\begin{eqnarray} 
\sigma(p\bar{p}\to W^*\!+X) & = & \sum_{i,j} \int dx_1 \int dx_2 \,
[q_i(x_1,\mu_F)\bar{q}_j(x_2,\mu_F)
+\bar{q}_i(x_1,\mu_F)q_j(x_2,\mu_F)] \\ & & \times |V_{ij}|^2
\frac{\pi^2\alpha_W}{3}\delta(x_1x_2S-q^2) \nonumber
\end{eqnarray} 
where $\alpha_W=g^2/4\pi \equiv \sqrt 2 G_{\mu}M_W^2/\pi$, $S$ is the
square of the total hadronic center-of-mass energy, $q$ and $\bar{q}$
are the parton distribution functions, $\mu_F$ is the factorization
scale, and the sum on $i$ and $j$ runs over all contributing
quark-antiquark combinations.  At leading order, the coefficient of
the $-g^{\mu\nu}$ term in the imaginary part of the $W$-boson
self-energy is
\begin{equation}
{\rm Im}\:\Pi(q^2,m_t^2,m_b^2) =
\frac{\alpha_W\lambda^{1/2}|V_{tb}|^2}{2} \left[ 1 -
\frac{m_t^2+m_b^2}{2q^2} - \frac{(m_t^2-m_b^2)^2}{2q^4} \right]
\end{equation}
where $\lambda$ is the triangle function associated with two-particle
phase space,
\begin{equation}
\lambda = \lambda(q^2,m_t^2,m_b^2) = q^4 + m_t^4 + m_b^4 - 2q^2m_t^2 -
2q^2m_b^2 - 2m_t^2m_b^2 \;.
\end{equation} 
Using Eq.~(1), the differential cross section is thus
\begin{eqnarray} 
\frac{d\sigma}{dq^2}(p\bar{p} \to t\bar{b} + X) & = & \sum_{i,j}\int
dx_1 \int dx_2 \, [q_i(x_1,\mu_F)\bar{q}_j(x_2,\mu_F)
+\bar{q}_i(x_1,\mu_F)q_j(x_2,\mu_F)] \\ & & \times |V_{ij}|^2
\frac{\pi\alpha_W^2\lambda^{1/2}|V_{tb}|^2} {12(q^2-M_W^2)^2} \left[ 1
- \frac{m_t^2+m_b^2}{2q^2} - \frac{(m_t^2-m_b^2)^2}{2q^4} \right]
\delta(x_1x_2S-q^2) \; . \nonumber
\label{LO}
\end{eqnarray} 
At leading order, the integration over $q^2$ to obtain the total cross
section is trivial due to the delta function.  At next-to-leading
order, however, it is necessary to perform the integration
numerically.

The ${\sl O}(\alpha_s)$ corrections to the Drell-Yan process
\cite{DY1} and the $W$-boson self energy \cite{STN,CGN} were both
calculated many years ago.  We use the expression for
$\sigma(p\bar{p}\to W^*+X)$ as given in Eqs.~(9.5) and (12.3) of
Ref.~\cite{TASI}, and Im $\Pi$ as derived from Eq.~(3.3) of
Ref.~\cite{CGN}.\footnote{The exact correspondence between our
notation and that of Ref.~\cite{CGN} is \mbox{Im $\Pi =
3\pi\alpha_W|V_{tb}|^2 {\rm Im}(\Pi_1^V+\Pi_1^A)$}.} We use $m_t$ =
175 GeV, $m_b$ = 5 GeV, $M_W$ = 80.33 GeV, $|V_{tb}|$ = 1, $G_{\mu} =
1.16639 \times 10^{-5}\;{\rm GeV}^{-2}$, and $\alpha_s$ as given by
the parton distribution functions.

\begin{figure}
\setlength{\unitlength}{0.240900pt}
\ifx\plotpoint\undefined\newsavebox{\plotpoint}\fi
\sbox{\plotpoint}{\rule[-0.200pt]{0.400pt}{0.400pt}}%
\begin{picture}(944,900)(0,0)
\font\gnuplot=cmr10 at 10pt
\gnuplot
\sbox{\plotpoint}{\rule[-0.200pt]{0.400pt}{0.400pt}}%
\put(220.0,113.0){\rule[-0.200pt]{4.818pt}{0.400pt}}
\put(198,113){\makebox(0,0)[r]{0.45}}
\put(860.0,113.0){\rule[-0.200pt]{4.818pt}{0.400pt}}
\put(220.0,185.0){\rule[-0.200pt]{4.818pt}{0.400pt}}
\put(198,185){\makebox(0,0)[r]{0.5}}
\put(860.0,185.0){\rule[-0.200pt]{4.818pt}{0.400pt}}
\put(220.0,257.0){\rule[-0.200pt]{4.818pt}{0.400pt}}
\put(198,257){\makebox(0,0)[r]{0.55}}
\put(860.0,257.0){\rule[-0.200pt]{4.818pt}{0.400pt}}
\put(220.0,329.0){\rule[-0.200pt]{4.818pt}{0.400pt}}
\put(198,329){\makebox(0,0)[r]{0.6}}
\put(860.0,329.0){\rule[-0.200pt]{4.818pt}{0.400pt}}
\put(220.0,401.0){\rule[-0.200pt]{4.818pt}{0.400pt}}
\put(198,401){\makebox(0,0)[r]{0.65}}
\put(860.0,401.0){\rule[-0.200pt]{4.818pt}{0.400pt}}
\put(220.0,473.0){\rule[-0.200pt]{4.818pt}{0.400pt}}
\put(198,473){\makebox(0,0)[r]{0.7}}
\put(860.0,473.0){\rule[-0.200pt]{4.818pt}{0.400pt}}
\put(220.0,544.0){\rule[-0.200pt]{4.818pt}{0.400pt}}
\put(198,544){\makebox(0,0)[r]{0.75}}
\put(860.0,544.0){\rule[-0.200pt]{4.818pt}{0.400pt}}
\put(220.0,616.0){\rule[-0.200pt]{4.818pt}{0.400pt}}
\put(198,616){\makebox(0,0)[r]{0.8}}
\put(860.0,616.0){\rule[-0.200pt]{4.818pt}{0.400pt}}
\put(220.0,688.0){\rule[-0.200pt]{4.818pt}{0.400pt}}
\put(198,688){\makebox(0,0)[r]{0.85}}
\put(860.0,688.0){\rule[-0.200pt]{4.818pt}{0.400pt}}
\put(220.0,760.0){\rule[-0.200pt]{4.818pt}{0.400pt}}
\put(198,760){\makebox(0,0)[r]{0.9}}
\put(860.0,760.0){\rule[-0.200pt]{4.818pt}{0.400pt}}
\put(220.0,832.0){\rule[-0.200pt]{4.818pt}{0.400pt}}
\put(198,832){\makebox(0,0)[r]{0.95}}
\put(860.0,832.0){\rule[-0.200pt]{4.818pt}{0.400pt}}
\put(220.0,113.0){\rule[-0.200pt]{0.400pt}{4.818pt}}
\put(220,68){\makebox(0,0){0.2}}
\put(220.0,812.0){\rule[-0.200pt]{0.400pt}{4.818pt}}
\put(303.0,113.0){\rule[-0.200pt]{0.400pt}{4.818pt}}
\put(303.0,812.0){\rule[-0.200pt]{0.400pt}{4.818pt}}
\put(362.0,113.0){\rule[-0.200pt]{0.400pt}{4.818pt}}
\put(362.0,812.0){\rule[-0.200pt]{0.400pt}{4.818pt}}
\put(408.0,113.0){\rule[-0.200pt]{0.400pt}{4.818pt}}
\put(408,68){\makebox(0,0){0.5}}
\put(408.0,812.0){\rule[-0.200pt]{0.400pt}{4.818pt}}
\put(445.0,113.0){\rule[-0.200pt]{0.400pt}{4.818pt}}
\put(445.0,812.0){\rule[-0.200pt]{0.400pt}{4.818pt}}
\put(477.0,113.0){\rule[-0.200pt]{0.400pt}{4.818pt}}
\put(477.0,812.0){\rule[-0.200pt]{0.400pt}{4.818pt}}
\put(504.0,113.0){\rule[-0.200pt]{0.400pt}{4.818pt}}
\put(504.0,812.0){\rule[-0.200pt]{0.400pt}{4.818pt}}
\put(528.0,113.0){\rule[-0.200pt]{0.400pt}{4.818pt}}
\put(528.0,812.0){\rule[-0.200pt]{0.400pt}{4.818pt}}
\put(550.0,113.0){\rule[-0.200pt]{0.400pt}{4.818pt}}
\put(550,68){\makebox(0,0){1}}
\put(550.0,812.0){\rule[-0.200pt]{0.400pt}{4.818pt}}
\put(692.0,113.0){\rule[-0.200pt]{0.400pt}{4.818pt}}
\put(692,68){\makebox(0,0){2}}
\put(692.0,812.0){\rule[-0.200pt]{0.400pt}{4.818pt}}
\put(775.0,113.0){\rule[-0.200pt]{0.400pt}{4.818pt}}
\put(775.0,812.0){\rule[-0.200pt]{0.400pt}{4.818pt}}
\put(834.0,113.0){\rule[-0.200pt]{0.400pt}{4.818pt}}
\put(834.0,812.0){\rule[-0.200pt]{0.400pt}{4.818pt}}
\put(880.0,113.0){\rule[-0.200pt]{0.400pt}{4.818pt}}
\put(880,68){\makebox(0,0){5}}
\put(880.0,812.0){\rule[-0.200pt]{0.400pt}{4.818pt}}
\put(220.0,113.0){\rule[-0.200pt]{158.994pt}{0.400pt}}
\put(880.0,113.0){\rule[-0.200pt]{0.400pt}{173.207pt}}
\put(220.0,832.0){\rule[-0.200pt]{158.994pt}{0.400pt}}
\put(67,472){\makebox(0,0){$\sigma$(pb)}}
\put(550,23){\makebox(0,0){$\mu_F/\sqrt{q^2}$}}
\put(550,877){\makebox(0,0){Factorization-scale dependence}}
\put(692,789){\makebox(0,0){\small Tevatron}}
\put(692,739){\makebox(0,0){\small $\sqrt{S}$ = 2 TeV}}
\put(550,645){\makebox(0,0){\small NLO (initial state)}}
\put(550,343){\makebox(0,0){\small LO}}
\put(220.0,113.0){\rule[-0.200pt]{0.400pt}{173.207pt}}
\multiput(220.00,452.92)(0.927,-0.492){21}{\rule{0.833pt}{0.119pt}}
\multiput(220.00,453.17)(20.270,-12.000){2}{\rule{0.417pt}{0.400pt}}
\multiput(242.00,440.92)(0.927,-0.495){35}{\rule{0.837pt}{0.119pt}}
\multiput(242.00,441.17)(33.263,-19.000){2}{\rule{0.418pt}{0.400pt}}
\multiput(277.00,421.92)(1.009,-0.495){31}{\rule{0.900pt}{0.119pt}}
\multiput(277.00,422.17)(32.132,-17.000){2}{\rule{0.450pt}{0.400pt}}
\multiput(311.00,404.92)(1.009,-0.495){31}{\rule{0.900pt}{0.119pt}}
\multiput(311.00,405.17)(32.132,-17.000){2}{\rule{0.450pt}{0.400pt}}
\multiput(345.00,387.92)(1.009,-0.495){31}{\rule{0.900pt}{0.119pt}}
\multiput(345.00,388.17)(32.132,-17.000){2}{\rule{0.450pt}{0.400pt}}
\multiput(379.00,370.92)(1.073,-0.494){29}{\rule{0.950pt}{0.119pt}}
\multiput(379.00,371.17)(32.028,-16.000){2}{\rule{0.475pt}{0.400pt}}
\multiput(413.00,354.92)(1.147,-0.494){27}{\rule{1.007pt}{0.119pt}}
\multiput(413.00,355.17)(31.911,-15.000){2}{\rule{0.503pt}{0.400pt}}
\multiput(447.00,339.92)(1.181,-0.494){27}{\rule{1.033pt}{0.119pt}}
\multiput(447.00,340.17)(32.855,-15.000){2}{\rule{0.517pt}{0.400pt}}
\multiput(482.00,324.92)(1.147,-0.494){27}{\rule{1.007pt}{0.119pt}}
\multiput(482.00,325.17)(31.911,-15.000){2}{\rule{0.503pt}{0.400pt}}
\multiput(516.00,309.92)(1.231,-0.494){25}{\rule{1.071pt}{0.119pt}}
\multiput(516.00,310.17)(31.776,-14.000){2}{\rule{0.536pt}{0.400pt}}
\multiput(550.00,295.92)(1.231,-0.494){25}{\rule{1.071pt}{0.119pt}}
\multiput(550.00,296.17)(31.776,-14.000){2}{\rule{0.536pt}{0.400pt}}
\multiput(584.00,281.92)(1.329,-0.493){23}{\rule{1.146pt}{0.119pt}}
\multiput(584.00,282.17)(31.621,-13.000){2}{\rule{0.573pt}{0.400pt}}
\multiput(618.00,268.92)(1.369,-0.493){23}{\rule{1.177pt}{0.119pt}}
\multiput(618.00,269.17)(32.557,-13.000){2}{\rule{0.588pt}{0.400pt}}
\multiput(653.00,255.92)(1.329,-0.493){23}{\rule{1.146pt}{0.119pt}}
\multiput(653.00,256.17)(31.621,-13.000){2}{\rule{0.573pt}{0.400pt}}
\multiput(687.00,242.92)(1.444,-0.492){21}{\rule{1.233pt}{0.119pt}}
\multiput(687.00,243.17)(31.440,-12.000){2}{\rule{0.617pt}{0.400pt}}
\multiput(721.00,230.92)(1.444,-0.492){21}{\rule{1.233pt}{0.119pt}}
\multiput(721.00,231.17)(31.440,-12.000){2}{\rule{0.617pt}{0.400pt}}
\multiput(755.00,218.92)(1.581,-0.492){19}{\rule{1.336pt}{0.118pt}}
\multiput(755.00,219.17)(31.226,-11.000){2}{\rule{0.668pt}{0.400pt}}
\multiput(789.00,207.92)(1.444,-0.492){21}{\rule{1.233pt}{0.119pt}}
\multiput(789.00,208.17)(31.440,-12.000){2}{\rule{0.617pt}{0.400pt}}
\multiput(823.00,195.92)(1.798,-0.491){17}{\rule{1.500pt}{0.118pt}}
\multiput(823.00,196.17)(31.887,-10.000){2}{\rule{0.750pt}{0.400pt}}
\multiput(858.00,185.93)(1.637,-0.485){11}{\rule{1.357pt}{0.117pt}}
\multiput(858.00,186.17)(19.183,-7.000){2}{\rule{0.679pt}{0.400pt}}
\multiput(220.00,651.94)(3.113,-0.468){5}{\rule{2.300pt}{0.113pt}}
\multiput(220.00,652.17)(17.226,-4.000){2}{\rule{1.150pt}{0.400pt}}
\multiput(242.00,647.93)(2.628,-0.485){11}{\rule{2.100pt}{0.117pt}}
\multiput(242.00,648.17)(30.641,-7.000){2}{\rule{1.050pt}{0.400pt}}
\multiput(277.00,640.93)(3.022,-0.482){9}{\rule{2.367pt}{0.116pt}}
\multiput(277.00,641.17)(29.088,-6.000){2}{\rule{1.183pt}{0.400pt}}
\multiput(311.00,634.93)(3.022,-0.482){9}{\rule{2.367pt}{0.116pt}}
\multiput(311.00,635.17)(29.088,-6.000){2}{\rule{1.183pt}{0.400pt}}
\multiput(345.00,628.93)(3.716,-0.477){7}{\rule{2.820pt}{0.115pt}}
\multiput(345.00,629.17)(28.147,-5.000){2}{\rule{1.410pt}{0.400pt}}
\multiput(379.00,623.93)(3.716,-0.477){7}{\rule{2.820pt}{0.115pt}}
\multiput(379.00,624.17)(28.147,-5.000){2}{\rule{1.410pt}{0.400pt}}
\multiput(413.00,618.93)(3.022,-0.482){9}{\rule{2.367pt}{0.116pt}}
\multiput(413.00,619.17)(29.088,-6.000){2}{\rule{1.183pt}{0.400pt}}
\multiput(447.00,612.94)(5.014,-0.468){5}{\rule{3.600pt}{0.113pt}}
\multiput(447.00,613.17)(27.528,-4.000){2}{\rule{1.800pt}{0.400pt}}
\multiput(482.00,608.93)(3.716,-0.477){7}{\rule{2.820pt}{0.115pt}}
\multiput(482.00,609.17)(28.147,-5.000){2}{\rule{1.410pt}{0.400pt}}
\multiput(516.00,603.93)(3.716,-0.477){7}{\rule{2.820pt}{0.115pt}}
\multiput(516.00,604.17)(28.147,-5.000){2}{\rule{1.410pt}{0.400pt}}
\multiput(550.00,598.94)(4.868,-0.468){5}{\rule{3.500pt}{0.113pt}}
\multiput(550.00,599.17)(26.736,-4.000){2}{\rule{1.750pt}{0.400pt}}
\multiput(584.00,594.93)(3.716,-0.477){7}{\rule{2.820pt}{0.115pt}}
\multiput(584.00,595.17)(28.147,-5.000){2}{\rule{1.410pt}{0.400pt}}
\multiput(618.00,589.94)(5.014,-0.468){5}{\rule{3.600pt}{0.113pt}}
\multiput(618.00,590.17)(27.528,-4.000){2}{\rule{1.800pt}{0.400pt}}
\multiput(653.00,585.94)(4.868,-0.468){5}{\rule{3.500pt}{0.113pt}}
\multiput(653.00,586.17)(26.736,-4.000){2}{\rule{1.750pt}{0.400pt}}
\multiput(687.00,581.94)(4.868,-0.468){5}{\rule{3.500pt}{0.113pt}}
\multiput(687.00,582.17)(26.736,-4.000){2}{\rule{1.750pt}{0.400pt}}
\multiput(721.00,577.94)(4.868,-0.468){5}{\rule{3.500pt}{0.113pt}}
\multiput(721.00,578.17)(26.736,-4.000){2}{\rule{1.750pt}{0.400pt}}
\multiput(755.00,573.94)(4.868,-0.468){5}{\rule{3.500pt}{0.113pt}}
\multiput(755.00,574.17)(26.736,-4.000){2}{\rule{1.750pt}{0.400pt}}
\multiput(789.00,569.95)(7.383,-0.447){3}{\rule{4.633pt}{0.108pt}}
\multiput(789.00,570.17)(24.383,-3.000){2}{\rule{2.317pt}{0.400pt}}
\multiput(823.00,566.95)(7.607,-0.447){3}{\rule{4.767pt}{0.108pt}}
\multiput(823.00,567.17)(25.107,-3.000){2}{\rule{2.383pt}{0.400pt}}
\put(858,563.67){\rule{5.300pt}{0.400pt}}
\multiput(858.00,564.17)(11.000,-1.000){2}{\rule{2.650pt}{0.400pt}}
\end{picture}
\setlength{\unitlength}{0.240900pt}
\ifx\plotpoint\undefined\newsavebox{\plotpoint}\fi
\begin{picture}(944,900)(0,0)
\font\gnuplot=cmr10 at 10pt
\gnuplot
\sbox{\plotpoint}{\rule[-0.200pt]{0.400pt}{0.400pt}}%
\put(220.0,113.0){\rule[-0.200pt]{4.818pt}{0.400pt}}
\put(198,113){\makebox(0,0)[r]{5.5}}
\put(860.0,113.0){\rule[-0.200pt]{4.818pt}{0.400pt}}
\put(220.0,185.0){\rule[-0.200pt]{4.818pt}{0.400pt}}
\put(198,185){\makebox(0,0)[r]{6}}
\put(860.0,185.0){\rule[-0.200pt]{4.818pt}{0.400pt}}
\put(220.0,257.0){\rule[-0.200pt]{4.818pt}{0.400pt}}
\put(198,257){\makebox(0,0)[r]{6.5}}
\put(860.0,257.0){\rule[-0.200pt]{4.818pt}{0.400pt}}
\put(220.0,329.0){\rule[-0.200pt]{4.818pt}{0.400pt}}
\put(198,329){\makebox(0,0)[r]{7}}
\put(860.0,329.0){\rule[-0.200pt]{4.818pt}{0.400pt}}
\put(220.0,401.0){\rule[-0.200pt]{4.818pt}{0.400pt}}
\put(198,401){\makebox(0,0)[r]{7.5}}
\put(860.0,401.0){\rule[-0.200pt]{4.818pt}{0.400pt}}
\put(220.0,473.0){\rule[-0.200pt]{4.818pt}{0.400pt}}
\put(198,473){\makebox(0,0)[r]{8}}
\put(860.0,473.0){\rule[-0.200pt]{4.818pt}{0.400pt}}
\put(220.0,544.0){\rule[-0.200pt]{4.818pt}{0.400pt}}
\put(198,544){\makebox(0,0)[r]{8.5}}
\put(860.0,544.0){\rule[-0.200pt]{4.818pt}{0.400pt}}
\put(220.0,616.0){\rule[-0.200pt]{4.818pt}{0.400pt}}
\put(198,616){\makebox(0,0)[r]{9}}
\put(860.0,616.0){\rule[-0.200pt]{4.818pt}{0.400pt}}
\put(220.0,688.0){\rule[-0.200pt]{4.818pt}{0.400pt}}
\put(198,688){\makebox(0,0)[r]{9.5}}
\put(860.0,688.0){\rule[-0.200pt]{4.818pt}{0.400pt}}
\put(220.0,760.0){\rule[-0.200pt]{4.818pt}{0.400pt}}
\put(198,760){\makebox(0,0)[r]{10}}
\put(860.0,760.0){\rule[-0.200pt]{4.818pt}{0.400pt}}
\put(220.0,832.0){\rule[-0.200pt]{4.818pt}{0.400pt}}
\put(198,832){\makebox(0,0)[r]{10.5}}
\put(860.0,832.0){\rule[-0.200pt]{4.818pt}{0.400pt}}
\put(220.0,113.0){\rule[-0.200pt]{0.400pt}{4.818pt}}
\put(220,68){\makebox(0,0){0.2}}
\put(220.0,812.0){\rule[-0.200pt]{0.400pt}{4.818pt}}
\put(303.0,113.0){\rule[-0.200pt]{0.400pt}{4.818pt}}
\put(303.0,812.0){\rule[-0.200pt]{0.400pt}{4.818pt}}
\put(362.0,113.0){\rule[-0.200pt]{0.400pt}{4.818pt}}
\put(362.0,812.0){\rule[-0.200pt]{0.400pt}{4.818pt}}
\put(408.0,113.0){\rule[-0.200pt]{0.400pt}{4.818pt}}
\put(408,68){\makebox(0,0){0.5}}
\put(408.0,812.0){\rule[-0.200pt]{0.400pt}{4.818pt}}
\put(445.0,113.0){\rule[-0.200pt]{0.400pt}{4.818pt}}
\put(445.0,812.0){\rule[-0.200pt]{0.400pt}{4.818pt}}
\put(477.0,113.0){\rule[-0.200pt]{0.400pt}{4.818pt}}
\put(477.0,812.0){\rule[-0.200pt]{0.400pt}{4.818pt}}
\put(504.0,113.0){\rule[-0.200pt]{0.400pt}{4.818pt}}
\put(504.0,812.0){\rule[-0.200pt]{0.400pt}{4.818pt}}
\put(528.0,113.0){\rule[-0.200pt]{0.400pt}{4.818pt}}
\put(528.0,812.0){\rule[-0.200pt]{0.400pt}{4.818pt}}
\put(550.0,113.0){\rule[-0.200pt]{0.400pt}{4.818pt}}
\put(550,68){\makebox(0,0){1}}
\put(550.0,812.0){\rule[-0.200pt]{0.400pt}{4.818pt}}
\put(692.0,113.0){\rule[-0.200pt]{0.400pt}{4.818pt}}
\put(692,68){\makebox(0,0){2}}
\put(692.0,812.0){\rule[-0.200pt]{0.400pt}{4.818pt}}
\put(775.0,113.0){\rule[-0.200pt]{0.400pt}{4.818pt}}
\put(775.0,812.0){\rule[-0.200pt]{0.400pt}{4.818pt}}
\put(834.0,113.0){\rule[-0.200pt]{0.400pt}{4.818pt}}
\put(834.0,812.0){\rule[-0.200pt]{0.400pt}{4.818pt}}
\put(880.0,113.0){\rule[-0.200pt]{0.400pt}{4.818pt}}
\put(880,68){\makebox(0,0){5}}
\put(880.0,812.0){\rule[-0.200pt]{0.400pt}{4.818pt}}
\put(220.0,113.0){\rule[-0.200pt]{158.994pt}{0.400pt}}
\put(880.0,113.0){\rule[-0.200pt]{0.400pt}{173.207pt}}
\put(220.0,832.0){\rule[-0.200pt]{158.994pt}{0.400pt}}
\put(89,472){\makebox(0,0){$\sigma$(pb)}}
\put(550,23){\makebox(0,0){$\mu_F/\sqrt{q^2}$}}
\put(550,877){\makebox(0,0){Factorization-scale dependence}}
\put(692,789){\makebox(0,0){\small LHC}}
\put(692,746){\makebox(0,0){\small $\sqrt{S}$ = 14 TeV}}
\put(550,674){\makebox(0,0){\small NLO (initial state)}}
\put(550,329){\makebox(0,0){\small LO}}
\put(220.0,113.0){\rule[-0.200pt]{0.400pt}{173.207pt}}
\multiput(220.00,178.59)(1.252,0.489){15}{\rule{1.078pt}{0.118pt}}
\multiput(220.00,177.17)(19.763,9.000){2}{\rule{0.539pt}{0.400pt}}
\multiput(242.00,187.58)(1.181,0.494){27}{\rule{1.033pt}{0.119pt}}
\multiput(242.00,186.17)(32.855,15.000){2}{\rule{0.517pt}{0.400pt}}
\multiput(277.00,202.58)(1.329,0.493){23}{\rule{1.146pt}{0.119pt}}
\multiput(277.00,201.17)(31.621,13.000){2}{\rule{0.573pt}{0.400pt}}
\multiput(311.00,215.58)(1.329,0.493){23}{\rule{1.146pt}{0.119pt}}
\multiput(311.00,214.17)(31.621,13.000){2}{\rule{0.573pt}{0.400pt}}
\multiput(345.00,228.58)(1.444,0.492){21}{\rule{1.233pt}{0.119pt}}
\multiput(345.00,227.17)(31.440,12.000){2}{\rule{0.617pt}{0.400pt}}
\multiput(379.00,240.58)(1.444,0.492){21}{\rule{1.233pt}{0.119pt}}
\multiput(379.00,239.17)(31.440,12.000){2}{\rule{0.617pt}{0.400pt}}
\multiput(413.00,252.58)(1.581,0.492){19}{\rule{1.336pt}{0.118pt}}
\multiput(413.00,251.17)(31.226,11.000){2}{\rule{0.668pt}{0.400pt}}
\multiput(447.00,263.58)(1.628,0.492){19}{\rule{1.373pt}{0.118pt}}
\multiput(447.00,262.17)(32.151,11.000){2}{\rule{0.686pt}{0.400pt}}
\multiput(482.00,274.58)(1.746,0.491){17}{\rule{1.460pt}{0.118pt}}
\multiput(482.00,273.17)(30.970,10.000){2}{\rule{0.730pt}{0.400pt}}
\multiput(516.00,284.59)(1.951,0.489){15}{\rule{1.611pt}{0.118pt}}
\multiput(516.00,283.17)(30.656,9.000){2}{\rule{0.806pt}{0.400pt}}
\multiput(550.00,293.59)(1.951,0.489){15}{\rule{1.611pt}{0.118pt}}
\multiput(550.00,292.17)(30.656,9.000){2}{\rule{0.806pt}{0.400pt}}
\multiput(584.00,302.59)(2.211,0.488){13}{\rule{1.800pt}{0.117pt}}
\multiput(584.00,301.17)(30.264,8.000){2}{\rule{0.900pt}{0.400pt}}
\multiput(618.00,310.59)(2.277,0.488){13}{\rule{1.850pt}{0.117pt}}
\multiput(618.00,309.17)(31.160,8.000){2}{\rule{0.925pt}{0.400pt}}
\multiput(653.00,318.59)(2.211,0.488){13}{\rule{1.800pt}{0.117pt}}
\multiput(653.00,317.17)(30.264,8.000){2}{\rule{0.900pt}{0.400pt}}
\multiput(687.00,326.59)(2.552,0.485){11}{\rule{2.043pt}{0.117pt}}
\multiput(687.00,325.17)(29.760,7.000){2}{\rule{1.021pt}{0.400pt}}
\multiput(721.00,333.59)(3.022,0.482){9}{\rule{2.367pt}{0.116pt}}
\multiput(721.00,332.17)(29.088,6.000){2}{\rule{1.183pt}{0.400pt}}
\multiput(755.00,339.59)(2.552,0.485){11}{\rule{2.043pt}{0.117pt}}
\multiput(755.00,338.17)(29.760,7.000){2}{\rule{1.021pt}{0.400pt}}
\multiput(789.00,346.59)(3.022,0.482){9}{\rule{2.367pt}{0.116pt}}
\multiput(789.00,345.17)(29.088,6.000){2}{\rule{1.183pt}{0.400pt}}
\multiput(823.00,352.59)(3.827,0.477){7}{\rule{2.900pt}{0.115pt}}
\multiput(823.00,351.17)(28.981,5.000){2}{\rule{1.450pt}{0.400pt}}
\multiput(858.00,357.60)(3.113,0.468){5}{\rule{2.300pt}{0.113pt}}
\multiput(858.00,356.17)(17.226,4.000){2}{\rule{1.150pt}{0.400pt}}
\multiput(220.00,557.59)(1.937,0.482){9}{\rule{1.567pt}{0.116pt}}
\multiput(220.00,556.17)(18.748,6.000){2}{\rule{0.783pt}{0.400pt}}
\multiput(242.00,563.59)(2.277,0.488){13}{\rule{1.850pt}{0.117pt}}
\multiput(242.00,562.17)(31.160,8.000){2}{\rule{0.925pt}{0.400pt}}
\multiput(277.00,571.59)(2.211,0.488){13}{\rule{1.800pt}{0.117pt}}
\multiput(277.00,570.17)(30.264,8.000){2}{\rule{0.900pt}{0.400pt}}
\multiput(311.00,579.59)(2.552,0.485){11}{\rule{2.043pt}{0.117pt}}
\multiput(311.00,578.17)(29.760,7.000){2}{\rule{1.021pt}{0.400pt}}
\multiput(345.00,586.59)(2.552,0.485){11}{\rule{2.043pt}{0.117pt}}
\multiput(345.00,585.17)(29.760,7.000){2}{\rule{1.021pt}{0.400pt}}
\multiput(379.00,593.59)(3.022,0.482){9}{\rule{2.367pt}{0.116pt}}
\multiput(379.00,592.17)(29.088,6.000){2}{\rule{1.183pt}{0.400pt}}
\multiput(413.00,599.59)(3.022,0.482){9}{\rule{2.367pt}{0.116pt}}
\multiput(413.00,598.17)(29.088,6.000){2}{\rule{1.183pt}{0.400pt}}
\multiput(447.00,605.59)(3.827,0.477){7}{\rule{2.900pt}{0.115pt}}
\multiput(447.00,604.17)(28.981,5.000){2}{\rule{1.450pt}{0.400pt}}
\multiput(482.00,610.59)(3.716,0.477){7}{\rule{2.820pt}{0.115pt}}
\multiput(482.00,609.17)(28.147,5.000){2}{\rule{1.410pt}{0.400pt}}
\multiput(516.00,615.59)(3.716,0.477){7}{\rule{2.820pt}{0.115pt}}
\multiput(516.00,614.17)(28.147,5.000){2}{\rule{1.410pt}{0.400pt}}
\multiput(550.00,620.60)(4.868,0.468){5}{\rule{3.500pt}{0.113pt}}
\multiput(550.00,619.17)(26.736,4.000){2}{\rule{1.750pt}{0.400pt}}
\multiput(584.00,624.59)(3.716,0.477){7}{\rule{2.820pt}{0.115pt}}
\multiput(584.00,623.17)(28.147,5.000){2}{\rule{1.410pt}{0.400pt}}
\multiput(618.00,629.60)(5.014,0.468){5}{\rule{3.600pt}{0.113pt}}
\multiput(618.00,628.17)(27.528,4.000){2}{\rule{1.800pt}{0.400pt}}
\multiput(653.00,633.60)(4.868,0.468){5}{\rule{3.500pt}{0.113pt}}
\multiput(653.00,632.17)(26.736,4.000){2}{\rule{1.750pt}{0.400pt}}
\multiput(687.00,637.60)(4.868,0.468){5}{\rule{3.500pt}{0.113pt}}
\multiput(687.00,636.17)(26.736,4.000){2}{\rule{1.750pt}{0.400pt}}
\multiput(721.00,641.60)(4.868,0.468){5}{\rule{3.500pt}{0.113pt}}
\multiput(721.00,640.17)(26.736,4.000){2}{\rule{1.750pt}{0.400pt}}
\multiput(755.00,645.60)(4.868,0.468){5}{\rule{3.500pt}{0.113pt}}
\multiput(755.00,644.17)(26.736,4.000){2}{\rule{1.750pt}{0.400pt}}
\multiput(789.00,649.60)(4.868,0.468){5}{\rule{3.500pt}{0.113pt}}
\multiput(789.00,648.17)(26.736,4.000){2}{\rule{1.750pt}{0.400pt}}
\multiput(823.00,653.59)(3.827,0.477){7}{\rule{2.900pt}{0.115pt}}
\multiput(823.00,652.17)(28.981,5.000){2}{\rule{1.450pt}{0.400pt}}
\multiput(858.00,658.61)(4.704,0.447){3}{\rule{3.033pt}{0.108pt}}
\multiput(858.00,657.17)(15.704,3.000){2}{\rule{1.517pt}{0.400pt}}
\end{picture}
\caption{\footnotesize Factorization-scale dependence of the
leading-order (LO) and next-to-leading-order (NLO) cross sections for
$q\bar q\to t\bar b, \bar t b$ at the Tevatron and the LHC.  The NLO
cross sections include only the initial-state QCD correction, and not
the final-state correction.  The LO cross sections are calculated with
the CTEQ3L LO parton distribution functions, and the NLO cross
sections with the CTEQ3M NLO parton distribution functions.}
\label{muf}
\end{figure}
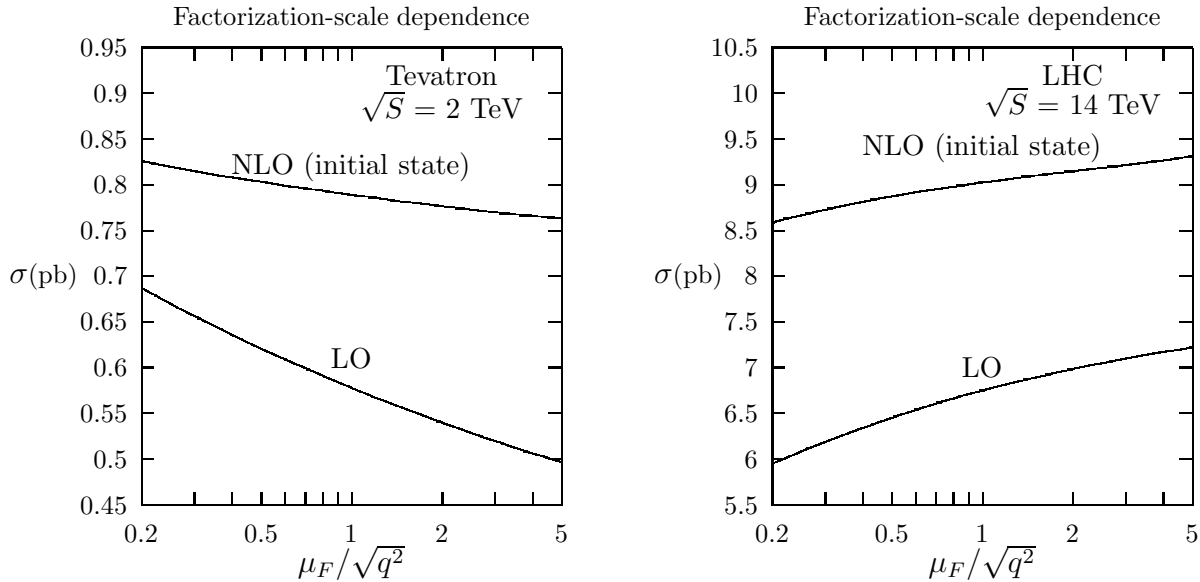

The calculation of the initial-state correction includes divergences
arising from collinear parton emission. These divergences cancel with
corresponding divergences present in the QCD correction to the parton
distribution functions.  The finite terms remaining depend on the
factorization scale $\mu_F$, both through the parton distribution
functions and explicitly in the partonic cross section.  The variation
of the leading-order and next-to-leading-order cross sections with
$\mu_F/\sqrt {q^2}$, where $\sqrt {q^2}$ is the mass of the virtual
$W$ boson,\footnote{We have chosen to refer the scale $\mu_F$ to the
$q^2$ of the virtual $W$ boson because this is the quantity which
appears in the factorization logarithms.  Thus the factorization scale
$\mu_F$ varies when integrating over $q^2$ to obtain the total cross
section.} is shown in Fig.~\ref{muf} at both the Tevatron and the
LHC. The leading-order cross section is calculated with the CTEQ3L
leading-order parton distribution functions, and the
next-to-leading-order cross section with the CTEQ3M
next-to-leading-order parton distribution functions \cite{CTEQ}.  The
leading-order cross section varies considerably with $\mu_F$, while
the next-to-leading-order cross section is appreciably less sensitive.
The next-to-leading-order cross section shown in Fig.~4 contains only
the initial-state correction.  We see that for $\mu_F=\sqrt {q^2}$ the
initial-state correction is +36\% at the Tevatron and +33\% at the
LHC.\footnote{If both the leading-order and next-to-leading-order
cross sections are calculated with the CTEQ3M next-to-leading-order
parton distribution functions, the initial-state correction is $+27\%$
at the Tevatron and $+15\%$ at the LHC.  Thus $+9\%$ of the
initial-state correction at the Tevatron, and $+18\%$ at the LHC, is
due to the increase in the leading-order cross section when it is
calculated with next-to-leading-order parton distribution functions.}
In what follows, we set $\mu_F=\sqrt{q^2}$.

\begin{figure}
\setlength{\unitlength}{0.240900pt}
\ifx\plotpoint\undefined\newsavebox{\plotpoint}\fi
\begin{picture}(944,900)(0,0)
\font\gnuplot=cmr10 at 10pt
\gnuplot
\sbox{\plotpoint}{\rule[-0.200pt]{0.400pt}{0.400pt}}%
\put(220.0,113.0){\rule[-0.200pt]{4.818pt}{0.400pt}}
\put(198,113){\makebox(0,0)[r]{0.5}}
\put(860.0,113.0){\rule[-0.200pt]{4.818pt}{0.400pt}}
\put(220.0,178.0){\rule[-0.200pt]{4.818pt}{0.400pt}}
\put(198,178){\makebox(0,0)[r]{0.55}}
\put(860.0,178.0){\rule[-0.200pt]{4.818pt}{0.400pt}}
\put(220.0,244.0){\rule[-0.200pt]{4.818pt}{0.400pt}}
\put(198,244){\makebox(0,0)[r]{0.6}}
\put(860.0,244.0){\rule[-0.200pt]{4.818pt}{0.400pt}}
\put(220.0,309.0){\rule[-0.200pt]{4.818pt}{0.400pt}}
\put(198,309){\makebox(0,0)[r]{0.65}}
\put(860.0,309.0){\rule[-0.200pt]{4.818pt}{0.400pt}}
\put(220.0,374.0){\rule[-0.200pt]{4.818pt}{0.400pt}}
\put(198,374){\makebox(0,0)[r]{0.7}}
\put(860.0,374.0){\rule[-0.200pt]{4.818pt}{0.400pt}}
\put(220.0,440.0){\rule[-0.200pt]{4.818pt}{0.400pt}}
\put(198,440){\makebox(0,0)[r]{0.75}}
\put(860.0,440.0){\rule[-0.200pt]{4.818pt}{0.400pt}}
\put(220.0,505.0){\rule[-0.200pt]{4.818pt}{0.400pt}}
\put(198,505){\makebox(0,0)[r]{0.8}}
\put(860.0,505.0){\rule[-0.200pt]{4.818pt}{0.400pt}}
\put(220.0,571.0){\rule[-0.200pt]{4.818pt}{0.400pt}}
\put(198,571){\makebox(0,0)[r]{0.85}}
\put(860.0,571.0){\rule[-0.200pt]{4.818pt}{0.400pt}}
\put(220.0,636.0){\rule[-0.200pt]{4.818pt}{0.400pt}}
\put(198,636){\makebox(0,0)[r]{0.9}}
\put(860.0,636.0){\rule[-0.200pt]{4.818pt}{0.400pt}}
\put(220.0,701.0){\rule[-0.200pt]{4.818pt}{0.400pt}}
\put(198,701){\makebox(0,0)[r]{0.95}}
\put(860.0,701.0){\rule[-0.200pt]{4.818pt}{0.400pt}}
\put(220.0,767.0){\rule[-0.200pt]{4.818pt}{0.400pt}}
\put(198,767){\makebox(0,0)[r]{1}}
\put(860.0,767.0){\rule[-0.200pt]{4.818pt}{0.400pt}}
\put(220.0,832.0){\rule[-0.200pt]{4.818pt}{0.400pt}}
\put(198,832){\makebox(0,0)[r]{1.05}}
\put(860.0,832.0){\rule[-0.200pt]{4.818pt}{0.400pt}}
\put(220.0,113.0){\rule[-0.200pt]{0.400pt}{4.818pt}}
\put(220,68){\makebox(0,0){0.2}}
\put(220.0,812.0){\rule[-0.200pt]{0.400pt}{4.818pt}}
\put(303.0,113.0){\rule[-0.200pt]{0.400pt}{4.818pt}}
\put(303.0,812.0){\rule[-0.200pt]{0.400pt}{4.818pt}}
\put(362.0,113.0){\rule[-0.200pt]{0.400pt}{4.818pt}}
\put(362.0,812.0){\rule[-0.200pt]{0.400pt}{4.818pt}}
\put(408.0,113.0){\rule[-0.200pt]{0.400pt}{4.818pt}}
\put(408,68){\makebox(0,0){0.5}}
\put(408.0,812.0){\rule[-0.200pt]{0.400pt}{4.818pt}}
\put(445.0,113.0){\rule[-0.200pt]{0.400pt}{4.818pt}}
\put(445.0,812.0){\rule[-0.200pt]{0.400pt}{4.818pt}}
\put(477.0,113.0){\rule[-0.200pt]{0.400pt}{4.818pt}}
\put(477.0,812.0){\rule[-0.200pt]{0.400pt}{4.818pt}}
\put(504.0,113.0){\rule[-0.200pt]{0.400pt}{4.818pt}}
\put(504.0,812.0){\rule[-0.200pt]{0.400pt}{4.818pt}}
\put(528.0,113.0){\rule[-0.200pt]{0.400pt}{4.818pt}}
\put(528.0,812.0){\rule[-0.200pt]{0.400pt}{4.818pt}}
\put(550.0,113.0){\rule[-0.200pt]{0.400pt}{4.818pt}}
\put(550,68){\makebox(0,0){1}}
\put(550.0,812.0){\rule[-0.200pt]{0.400pt}{4.818pt}}
\put(692.0,113.0){\rule[-0.200pt]{0.400pt}{4.818pt}}
\put(692,68){\makebox(0,0){2}}
\put(692.0,812.0){\rule[-0.200pt]{0.400pt}{4.818pt}}
\put(775.0,113.0){\rule[-0.200pt]{0.400pt}{4.818pt}}
\put(775.0,812.0){\rule[-0.200pt]{0.400pt}{4.818pt}}
\put(834.0,113.0){\rule[-0.200pt]{0.400pt}{4.818pt}}
\put(834.0,812.0){\rule[-0.200pt]{0.400pt}{4.818pt}}
\put(880.0,113.0){\rule[-0.200pt]{0.400pt}{4.818pt}}
\put(880,68){\makebox(0,0){5}}
\put(880.0,812.0){\rule[-0.200pt]{0.400pt}{4.818pt}}
\put(220.0,113.0){\rule[-0.200pt]{158.994pt}{0.400pt}}
\put(880.0,113.0){\rule[-0.200pt]{0.400pt}{173.207pt}}
\put(220.0,832.0){\rule[-0.200pt]{158.994pt}{0.400pt}}
\put(67,472){\makebox(0,0){$\sigma$(pb)}}
\put(550,23){\makebox(0,0){$\mu_R/\sqrt{q^2}$}}
\put(550,877){\makebox(0,0){Renormalization-scale dependence}}
\put(692,780){\makebox(0,0){\small Tevatron}}
\put(692,727){\makebox(0,0){\small $\sqrt{S}$ = 2 TeV}}
\put(550,662){\makebox(0,0){\small NLO}}
\put(550,244){\makebox(0,0){\small LO}}
\put(220.0,113.0){\rule[-0.200pt]{0.400pt}{173.207pt}}
\put(220.0,215.0){\rule[-0.200pt]{158.994pt}{0.400pt}}
\multiput(220.00,710.93)(1.418,-0.488){13}{\rule{1.200pt}{0.117pt}}
\multiput(220.00,711.17)(19.509,-8.000){2}{\rule{0.600pt}{0.400pt}}
\multiput(242.00,702.92)(1.798,-0.491){17}{\rule{1.500pt}{0.118pt}}
\multiput(242.00,703.17)(31.887,-10.000){2}{\rule{0.750pt}{0.400pt}}
\multiput(277.00,692.92)(1.746,-0.491){17}{\rule{1.460pt}{0.118pt}}
\multiput(277.00,693.17)(30.970,-10.000){2}{\rule{0.730pt}{0.400pt}}
\multiput(311.00,682.92)(1.746,-0.491){17}{\rule{1.460pt}{0.118pt}}
\multiput(311.00,683.17)(30.970,-10.000){2}{\rule{0.730pt}{0.400pt}}
\multiput(345.00,672.93)(1.951,-0.489){15}{\rule{1.611pt}{0.118pt}}
\multiput(345.00,673.17)(30.656,-9.000){2}{\rule{0.806pt}{0.400pt}}
\multiput(379.00,663.93)(2.211,-0.488){13}{\rule{1.800pt}{0.117pt}}
\multiput(379.00,664.17)(30.264,-8.000){2}{\rule{0.900pt}{0.400pt}}
\multiput(413.00,655.93)(2.211,-0.488){13}{\rule{1.800pt}{0.117pt}}
\multiput(413.00,656.17)(30.264,-8.000){2}{\rule{0.900pt}{0.400pt}}
\multiput(447.00,647.93)(2.277,-0.488){13}{\rule{1.850pt}{0.117pt}}
\multiput(447.00,648.17)(31.160,-8.000){2}{\rule{0.925pt}{0.400pt}}
\multiput(482.00,639.93)(2.552,-0.485){11}{\rule{2.043pt}{0.117pt}}
\multiput(482.00,640.17)(29.760,-7.000){2}{\rule{1.021pt}{0.400pt}}
\multiput(516.00,632.93)(2.552,-0.485){11}{\rule{2.043pt}{0.117pt}}
\multiput(516.00,633.17)(29.760,-7.000){2}{\rule{1.021pt}{0.400pt}}
\multiput(550.00,625.93)(3.022,-0.482){9}{\rule{2.367pt}{0.116pt}}
\multiput(550.00,626.17)(29.088,-6.000){2}{\rule{1.183pt}{0.400pt}}
\multiput(584.00,619.93)(3.022,-0.482){9}{\rule{2.367pt}{0.116pt}}
\multiput(584.00,620.17)(29.088,-6.000){2}{\rule{1.183pt}{0.400pt}}
\multiput(618.00,613.93)(3.112,-0.482){9}{\rule{2.433pt}{0.116pt}}
\multiput(618.00,614.17)(29.949,-6.000){2}{\rule{1.217pt}{0.400pt}}
\multiput(653.00,607.93)(3.022,-0.482){9}{\rule{2.367pt}{0.116pt}}
\multiput(653.00,608.17)(29.088,-6.000){2}{\rule{1.183pt}{0.400pt}}
\multiput(687.00,601.93)(3.716,-0.477){7}{\rule{2.820pt}{0.115pt}}
\multiput(687.00,602.17)(28.147,-5.000){2}{\rule{1.410pt}{0.400pt}}
\multiput(721.00,596.93)(3.716,-0.477){7}{\rule{2.820pt}{0.115pt}}
\multiput(721.00,597.17)(28.147,-5.000){2}{\rule{1.410pt}{0.400pt}}
\multiput(755.00,591.93)(3.716,-0.477){7}{\rule{2.820pt}{0.115pt}}
\multiput(755.00,592.17)(28.147,-5.000){2}{\rule{1.410pt}{0.400pt}}
\multiput(789.00,586.93)(3.716,-0.477){7}{\rule{2.820pt}{0.115pt}}
\multiput(789.00,587.17)(28.147,-5.000){2}{\rule{1.410pt}{0.400pt}}
\multiput(823.00,581.93)(3.827,-0.477){7}{\rule{2.900pt}{0.115pt}}
\multiput(823.00,582.17)(28.981,-5.000){2}{\rule{1.450pt}{0.400pt}}
\multiput(858.00,576.95)(4.704,-0.447){3}{\rule{3.033pt}{0.108pt}}
\multiput(858.00,577.17)(15.704,-3.000){2}{\rule{1.517pt}{0.400pt}}
\end{picture}
\setlength{\unitlength}{0.240900pt}
\ifx\plotpoint\undefined\newsavebox{\plotpoint}\fi
\begin{picture}(944,900)(0,0)
\font\gnuplot=cmr10 at 10pt
\gnuplot
\sbox{\plotpoint}{\rule[-0.200pt]{0.400pt}{0.400pt}}%
\put(220.0,113.0){\rule[-0.200pt]{4.818pt}{0.400pt}}
\put(198,113){\makebox(0,0)[r]{6}}
\put(860.0,113.0){\rule[-0.200pt]{4.818pt}{0.400pt}}
\put(220.0,173.0){\rule[-0.200pt]{4.818pt}{0.400pt}}
\put(198,173){\makebox(0,0)[r]{6.5}}
\put(860.0,173.0){\rule[-0.200pt]{4.818pt}{0.400pt}}
\put(220.0,233.0){\rule[-0.200pt]{4.818pt}{0.400pt}}
\put(198,233){\makebox(0,0)[r]{7}}
\put(860.0,233.0){\rule[-0.200pt]{4.818pt}{0.400pt}}
\put(220.0,293.0){\rule[-0.200pt]{4.818pt}{0.400pt}}
\put(198,293){\makebox(0,0)[r]{7.5}}
\put(860.0,293.0){\rule[-0.200pt]{4.818pt}{0.400pt}}
\put(220.0,353.0){\rule[-0.200pt]{4.818pt}{0.400pt}}
\put(198,353){\makebox(0,0)[r]{8}}
\put(860.0,353.0){\rule[-0.200pt]{4.818pt}{0.400pt}}
\put(220.0,413.0){\rule[-0.200pt]{4.818pt}{0.400pt}}
\put(198,413){\makebox(0,0)[r]{8.5}}
\put(860.0,413.0){\rule[-0.200pt]{4.818pt}{0.400pt}}
\put(220.0,473.0){\rule[-0.200pt]{4.818pt}{0.400pt}}
\put(198,473){\makebox(0,0)[r]{9}}
\put(860.0,473.0){\rule[-0.200pt]{4.818pt}{0.400pt}}
\put(220.0,532.0){\rule[-0.200pt]{4.818pt}{0.400pt}}
\put(198,532){\makebox(0,0)[r]{9.5}}
\put(860.0,532.0){\rule[-0.200pt]{4.818pt}{0.400pt}}
\put(220.0,592.0){\rule[-0.200pt]{4.818pt}{0.400pt}}
\put(198,592){\makebox(0,0)[r]{10}}
\put(860.0,592.0){\rule[-0.200pt]{4.818pt}{0.400pt}}
\put(220.0,652.0){\rule[-0.200pt]{4.818pt}{0.400pt}}
\put(198,652){\makebox(0,0)[r]{10.5}}
\put(860.0,652.0){\rule[-0.200pt]{4.818pt}{0.400pt}}
\put(220.0,712.0){\rule[-0.200pt]{4.818pt}{0.400pt}}
\put(198,712){\makebox(0,0)[r]{11}}
\put(860.0,712.0){\rule[-0.200pt]{4.818pt}{0.400pt}}
\put(220.0,772.0){\rule[-0.200pt]{4.818pt}{0.400pt}}
\put(198,772){\makebox(0,0)[r]{11.5}}
\put(860.0,772.0){\rule[-0.200pt]{4.818pt}{0.400pt}}
\put(220.0,832.0){\rule[-0.200pt]{4.818pt}{0.400pt}}
\put(198,832){\makebox(0,0)[r]{12}}
\put(860.0,832.0){\rule[-0.200pt]{4.818pt}{0.400pt}}
\put(220.0,113.0){\rule[-0.200pt]{0.400pt}{4.818pt}}
\put(220,68){\makebox(0,0){0.2}}
\put(220.0,812.0){\rule[-0.200pt]{0.400pt}{4.818pt}}
\put(303.0,113.0){\rule[-0.200pt]{0.400pt}{4.818pt}}
\put(303.0,812.0){\rule[-0.200pt]{0.400pt}{4.818pt}}
\put(362.0,113.0){\rule[-0.200pt]{0.400pt}{4.818pt}}
\put(362.0,812.0){\rule[-0.200pt]{0.400pt}{4.818pt}}
\put(408.0,113.0){\rule[-0.200pt]{0.400pt}{4.818pt}}
\put(408,68){\makebox(0,0){0.5}}
\put(408.0,812.0){\rule[-0.200pt]{0.400pt}{4.818pt}}
\put(445.0,113.0){\rule[-0.200pt]{0.400pt}{4.818pt}}
\put(445.0,812.0){\rule[-0.200pt]{0.400pt}{4.818pt}}
\put(477.0,113.0){\rule[-0.200pt]{0.400pt}{4.818pt}}
\put(477.0,812.0){\rule[-0.200pt]{0.400pt}{4.818pt}}
\put(504.0,113.0){\rule[-0.200pt]{0.400pt}{4.818pt}}
\put(504.0,812.0){\rule[-0.200pt]{0.400pt}{4.818pt}}
\put(528.0,113.0){\rule[-0.200pt]{0.400pt}{4.818pt}}
\put(528.0,812.0){\rule[-0.200pt]{0.400pt}{4.818pt}}
\put(550.0,113.0){\rule[-0.200pt]{0.400pt}{4.818pt}}
\put(550,68){\makebox(0,0){1}}
\put(550.0,812.0){\rule[-0.200pt]{0.400pt}{4.818pt}}
\put(692.0,113.0){\rule[-0.200pt]{0.400pt}{4.818pt}}
\put(692,68){\makebox(0,0){2}}
\put(692.0,812.0){\rule[-0.200pt]{0.400pt}{4.818pt}}
\put(775.0,113.0){\rule[-0.200pt]{0.400pt}{4.818pt}}
\put(775.0,812.0){\rule[-0.200pt]{0.400pt}{4.818pt}}
\put(834.0,113.0){\rule[-0.200pt]{0.400pt}{4.818pt}}
\put(834.0,812.0){\rule[-0.200pt]{0.400pt}{4.818pt}}
\put(880.0,113.0){\rule[-0.200pt]{0.400pt}{4.818pt}}
\put(880,68){\makebox(0,0){5}}
\put(880.0,812.0){\rule[-0.200pt]{0.400pt}{4.818pt}}
\put(220.0,113.0){\rule[-0.200pt]{158.994pt}{0.400pt}}
\put(880.0,113.0){\rule[-0.200pt]{0.400pt}{173.207pt}}
\put(220.0,832.0){\rule[-0.200pt]{158.994pt}{0.400pt}}
\put(89,472){\makebox(0,0){$\sigma$(pb)}}
\put(550,23){\makebox(0,0){$\mu_R/\sqrt{q^2}$}}
\put(550,877){\makebox(0,0){Renormalization-scale dependence}}
\put(692,784){\makebox(0,0){\small LHC}}
\put(692,736){\makebox(0,0){\small $\sqrt{S}$ = 14 TeV}}
\put(550,652){\makebox(0,0){\small NLO}}
\put(550,233){\makebox(0,0){\small LO}}
\put(220.0,113.0){\rule[-0.200pt]{0.400pt}{173.207pt}}
\put(220.0,203.0){\rule[-0.200pt]{158.994pt}{0.400pt}}
\multiput(220.00,682.93)(1.937,-0.482){9}{\rule{1.567pt}{0.116pt}}
\multiput(220.00,683.17)(18.748,-6.000){2}{\rule{0.783pt}{0.400pt}}
\multiput(242.00,676.93)(2.009,-0.489){15}{\rule{1.656pt}{0.118pt}}
\multiput(242.00,677.17)(31.564,-9.000){2}{\rule{0.828pt}{0.400pt}}
\multiput(277.00,667.93)(2.211,-0.488){13}{\rule{1.800pt}{0.117pt}}
\multiput(277.00,668.17)(30.264,-8.000){2}{\rule{0.900pt}{0.400pt}}
\multiput(311.00,659.93)(2.552,-0.485){11}{\rule{2.043pt}{0.117pt}}
\multiput(311.00,660.17)(29.760,-7.000){2}{\rule{1.021pt}{0.400pt}}
\multiput(345.00,652.93)(2.211,-0.488){13}{\rule{1.800pt}{0.117pt}}
\multiput(345.00,653.17)(30.264,-8.000){2}{\rule{0.900pt}{0.400pt}}
\multiput(379.00,644.93)(3.022,-0.482){9}{\rule{2.367pt}{0.116pt}}
\multiput(379.00,645.17)(29.088,-6.000){2}{\rule{1.183pt}{0.400pt}}
\multiput(413.00,638.93)(2.552,-0.485){11}{\rule{2.043pt}{0.117pt}}
\multiput(413.00,639.17)(29.760,-7.000){2}{\rule{1.021pt}{0.400pt}}
\multiput(447.00,631.93)(3.112,-0.482){9}{\rule{2.433pt}{0.116pt}}
\multiput(447.00,632.17)(29.949,-6.000){2}{\rule{1.217pt}{0.400pt}}
\multiput(482.00,625.93)(3.022,-0.482){9}{\rule{2.367pt}{0.116pt}}
\multiput(482.00,626.17)(29.088,-6.000){2}{\rule{1.183pt}{0.400pt}}
\multiput(516.00,619.93)(3.716,-0.477){7}{\rule{2.820pt}{0.115pt}}
\multiput(516.00,620.17)(28.147,-5.000){2}{\rule{1.410pt}{0.400pt}}
\multiput(550.00,614.93)(3.716,-0.477){7}{\rule{2.820pt}{0.115pt}}
\multiput(550.00,615.17)(28.147,-5.000){2}{\rule{1.410pt}{0.400pt}}
\multiput(584.00,609.93)(3.716,-0.477){7}{\rule{2.820pt}{0.115pt}}
\multiput(584.00,610.17)(28.147,-5.000){2}{\rule{1.410pt}{0.400pt}}
\multiput(618.00,604.93)(3.827,-0.477){7}{\rule{2.900pt}{0.115pt}}
\multiput(618.00,605.17)(28.981,-5.000){2}{\rule{1.450pt}{0.400pt}}
\multiput(653.00,599.93)(3.716,-0.477){7}{\rule{2.820pt}{0.115pt}}
\multiput(653.00,600.17)(28.147,-5.000){2}{\rule{1.410pt}{0.400pt}}
\multiput(687.00,594.94)(4.868,-0.468){5}{\rule{3.500pt}{0.113pt}}
\multiput(687.00,595.17)(26.736,-4.000){2}{\rule{1.750pt}{0.400pt}}
\multiput(721.00,590.93)(3.716,-0.477){7}{\rule{2.820pt}{0.115pt}}
\multiput(721.00,591.17)(28.147,-5.000){2}{\rule{1.410pt}{0.400pt}}
\multiput(755.00,585.94)(4.868,-0.468){5}{\rule{3.500pt}{0.113pt}}
\multiput(755.00,586.17)(26.736,-4.000){2}{\rule{1.750pt}{0.400pt}}
\multiput(789.00,581.94)(4.868,-0.468){5}{\rule{3.500pt}{0.113pt}}
\multiput(789.00,582.17)(26.736,-4.000){2}{\rule{1.750pt}{0.400pt}}
\multiput(823.00,577.94)(5.014,-0.468){5}{\rule{3.600pt}{0.113pt}}
\multiput(823.00,578.17)(27.528,-4.000){2}{\rule{1.800pt}{0.400pt}}
\put(858,573.17){\rule{4.500pt}{0.400pt}}
\multiput(858.00,574.17)(12.660,-2.000){2}{\rule{2.250pt}{0.400pt}}
\end{picture}
\caption{\footnotesize Renormalization-scale dependence of the
leading-order (LO) and next-to-leading-order (NLO) cross sections for
$q\bar q\to t\bar b, \bar t b$ at the Tevatron and the LHC.  The NLO
cross sections include both the initial-state and final-state
correction.  The LO cross sections are calculated with the CTEQ3L LO
parton distribution functions, and the NLO cross sections with the
CTEQ3M NLO parton distribution functions.}
\label{mur}
\end{figure}
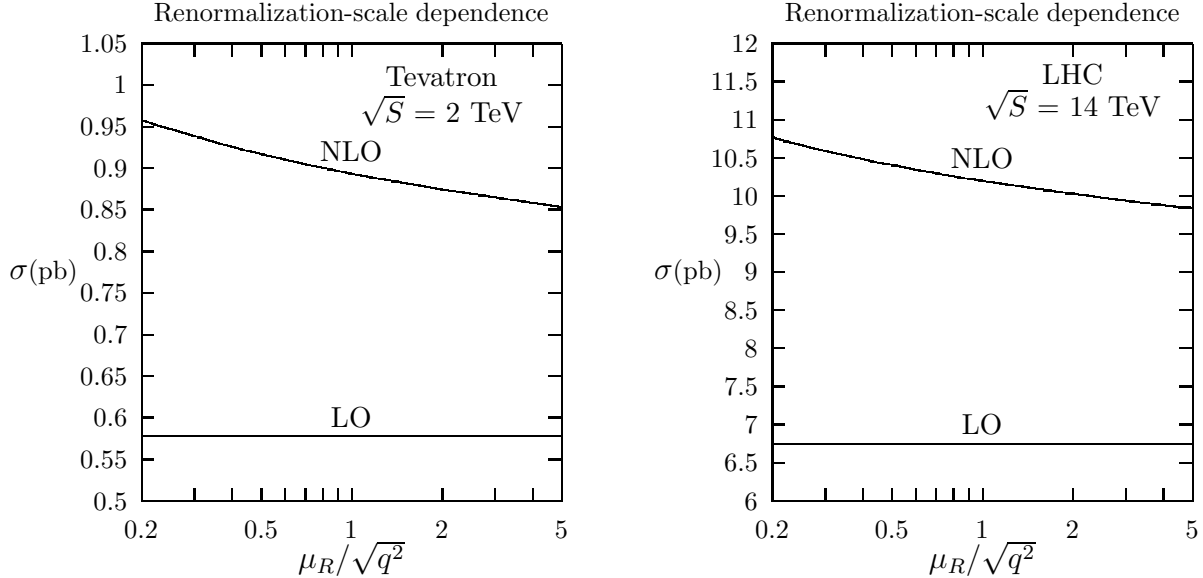

The cross section at next-to-leading order also depends on the
renormalization scale, $\mu_R$, at which $\alpha_s$ is evaluated. In
Fig.~\ref{mur} we show the next-to-leading-order cross section,
including both initial- and final-state corrections, as a function of
$\mu_R/\sqrt {q^2}$, at both the Tevatron and the LHC.  The dependence
of the cross section on the renormalization scale first appears at
next-to-leading order and is therefore mild. In what follows, we set
$\mu_R=\sqrt{q^2}$. The final-state correction increases the cross
section by $+18\%$ of the leading-order cross section at the Tevatron
and $+17\%$ at the LHC.

\begin{figure}
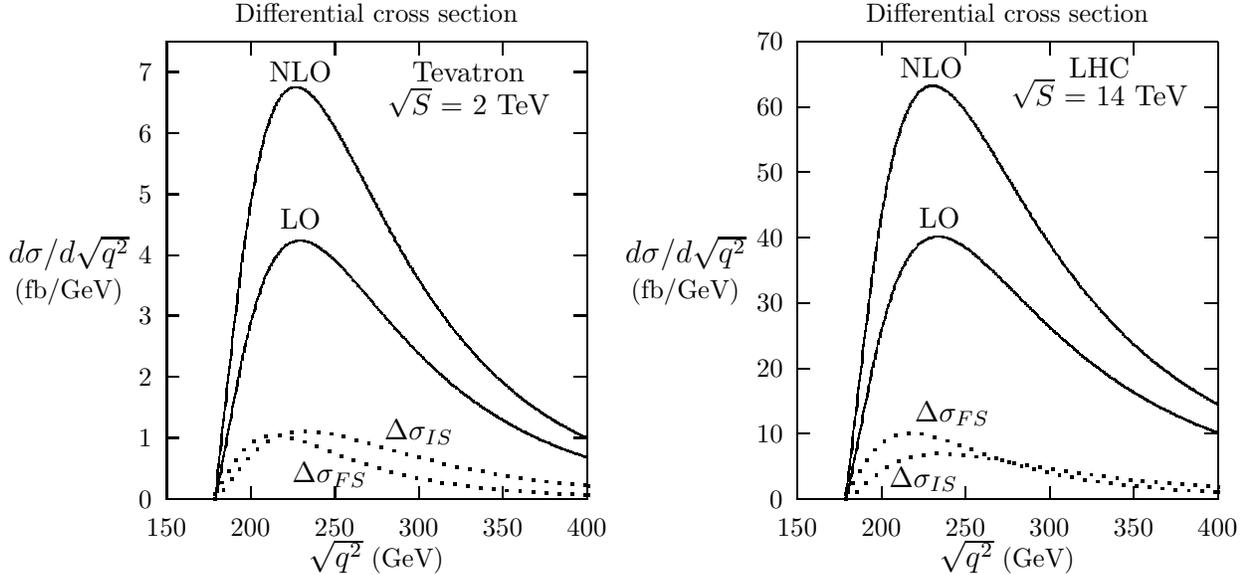

\input{dsig_tev.tex}
\input{dsig_lhc.tex}
\caption{\footnotesize Differential cross section for $q\bar q\to
t\bar b,\bar t b$ versus the mass of the virtual $s$-channel $W$
boson, at the Tevatron and the LHC. Both the leading-order (LO) and
next-to-leading order (NLO) cross sections are shown, as well as the
separate contributions from the initial-state (IS) and final-state
(FS) corrections.  The LO cross sections are calculated with the
CTEQ3L LO parton distribution functions, and the NLO cross sections
with the CTEQ3M NLO parton distribution functions.}
\label{dsig}
\end{figure}

We show in Fig.~\ref{dsig} the leading-order and next-to-leading-order
differential cross section as a function of the mass of the virtual
$W$ boson, $\sqrt {q^2}$, at both the Tevatron and the LHC.  Also
shown are the separate ${\sl O}(\alpha_s)$ corrections from the
initial and final states. These corrections have different shapes from
the leading-order cross section, and from each other.  In order to
observe $t\bar b$ production experimentally, it is necessary to detect
the $\bar b$ quark \cite{SW}.  Thus the measured cross section will
exclude some region near threshold, where the $\bar b$ quark does not
have sufficient transverse momentum to be detected with high
efficiency.  Therefore the measured cross section, as well as the QCD
correction, will depend on the acceptance for the $\bar b$ quark.

If the top and bottom quarks were stable, they would form quarkonium
bound states just below threshold \cite{CPras}.  We estimate the
distance below threshold that the ground state would occur, by analogy
with the hydrogen atom, to be $E \approx (4\alpha_s/3)^2 m_b/2 \approx
50$ MeV.\footnote{Here $m_b$ is the approximate reduced mass of the
system, and $C_F = 4/3$ is the usual SU(3) group theory factor
associated with the fundamental representation.} The formation time of
the ground state is approximately $1/E$. This is much greater than the
top-quark lifetime, $\Gamma_t^{-1} \approx (1.5\:{\rm GeV})^{-1}$, so
there is not sufficient time for quarkonium bound states to form
\cite{FK}.

Because the top-quark width is small compared to its mass,
interference between the corrections to production and decay
amplitudes has a negligible effect, of order $\alpha_s\Gamma_t/m_t$,
on the total cross section \cite{FKM}.  This interference also has a
negligible effect on differential cross sections, such as the
invariant-mass distribution of the decay products of the top quark
\cite{P}.

Our final results for the cross section and uncertainty will be
presented in Section 4.

\section{Yukawa correction}

\begin{figure}
\centerline{\psfig{figure=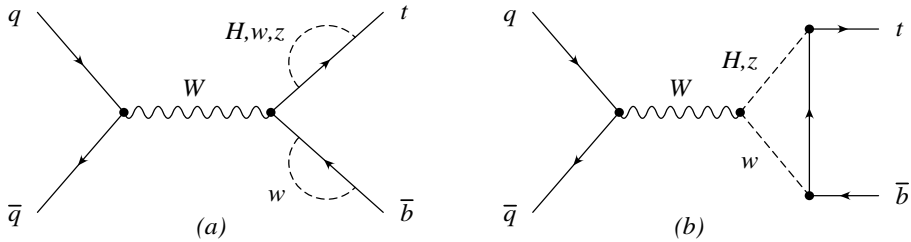,width=5in}}
\caption{\footnotesize ${\sl O}(\alpha_W m_t^2/M_W^2)$ corrections to
$q \bar q \to t \bar b$.  The dashed lines represent the Higgs boson
and the unphysical scalar $W$ and $Z$ bosons (in $R_{\xi}$ gauge): (a)
wavefunction renormalization, (b) vertex correction.}
\label{Yukawa graphs}
\end{figure}

\indent\indent The diagrams which contribute to the ${\sl O}(\alpha_W
m_t^2/m_W^2)$ Yukawa correction to $q \bar q \to t \bar b$ are shown
in Fig.~\ref{Yukawa graphs}.  The dashed lines represent the Higgs
boson and the unphysical scalar $W$ and $Z$ bosons associated with the
Higgs field (in the $R_{\xi}$ gauge).  The effect of a top-quark loop
in the $W$-boson propagator, which might be expected to contribute a
term of Yukawa strength, is absorbed by the renormalized weak coupling
constant, which we express in terms of $G_{\mu}$, the Fermi constant
measured in muon decay ($\alpha_W=g^2/4\pi \equiv \sqrt 2
G_{\mu}M_W^2/\pi$). We use standard Feynman integral techniques with
dimensional regularization to calculate the loop diagrams \cite{PV}, and
work in the approximation where the bottom quark is massless.  
Our other parameters are $m_t$ = 175 GeV, $M_W$ = 80.33 GeV,
$|V_{tb}|$ = 1, and $G_{\mu} = 1.16639 \times 10^{-5}\;{\rm
GeV}^{-2}$.

In the $m_b=0$ approximation, the matrix element of the $t \bar b$
current may be written as
\begin{eqnarray}
i \bar{u}(p_t)\Gamma^{\mu A} v(p_b) & = & \left(
\frac{-igT^A}{2\sqrt{2}} \right) \left\{ \bar{u}(p_t)
\gamma^{\mu}(1-\gamma^5) v(p_b) \frac{}{} \right. \\ & & \left. +
\left( \frac{m_t^2 G_{\mu}}{8\sqrt{2}\pi^2} \right) \bar{u}(p_t)
\left[ \gamma^{\mu} F_1(q^2) + \frac{(p_t^{\mu} - p_b^{\mu})}{m_t}
F_2(q^2) \right] (1-\gamma^5) v(p_b) \right\} \nonumber
\end{eqnarray}
where $p_t$ and $p_b$ are the outgoing four-momenta of the $t$ and
$\bar b$ quarks, respectively; the form factors $F_1$ and $F_2$ are
functions of $q^2=(p_t+p_b)^2$ and the Higgs-boson mass; and $T^A$ is
an SU(3) matrix [Tr$(T^A T^B)=\frac{1}{2} \delta^{AB}$].  The
fractional change in the differential cross section as a function of
the $q^2$ of the virtual $W$ boson is
\begin{equation}
\frac{\Delta d\sigma_Y/d\sqrt{q^2}}{d\sigma_{LO}/d\sqrt{q^2}} =
\left(\frac{m_t^2 G_{\mu}}{8 \sqrt 2 \pi^2} \right) \left[ 2F_1(q^2) +
F_2(q^2) \frac{(q^2 - m_t^2)^2} {q^4 -\frac{1}{2} q^2 m_t^2 -
\frac{1}{2} m_t^4} \right] \;.
\end{equation}
Analytic expressions for the form factors $F_1$ and $F_2$ are given in
an appendix.

\begin{figure}
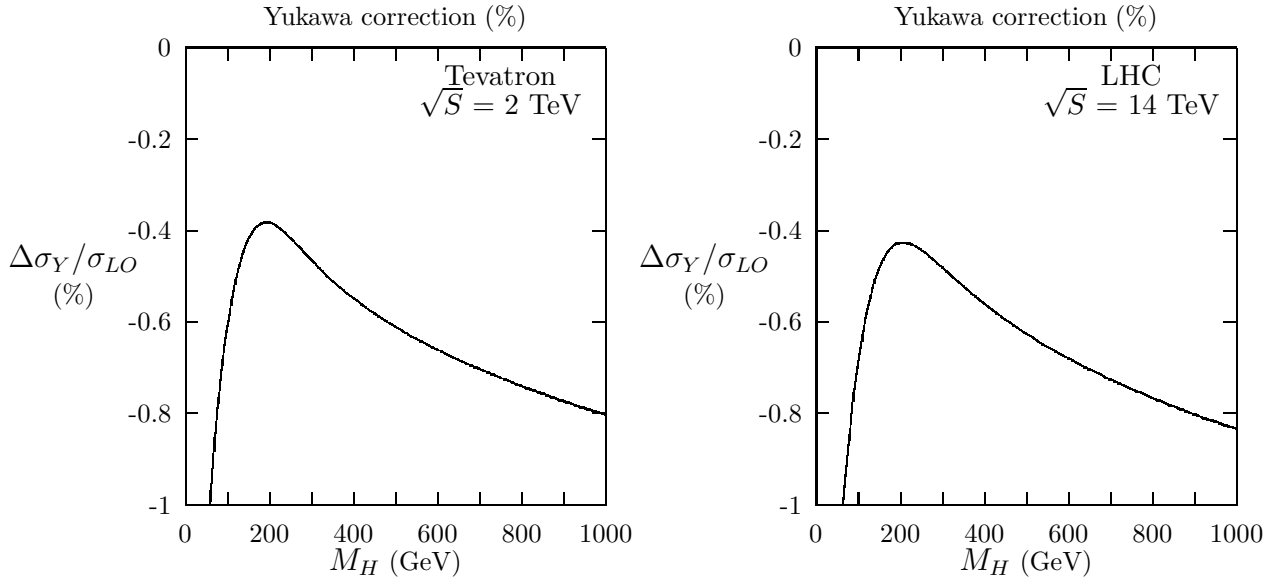

\input{yuk_tev.tex}
\input{yuk_lhc.tex}
\caption{\footnotesize Fractional change in the total cross section
for $q\bar q\to t\bar b, \bar t b$ due to the Yukawa correction
vs.~the Higgs-boson mass at the Tevatron and the LHC.}
\label{Yuk}
\end{figure}

The fractional change in the total cross section,
$\Delta\sigma_Y/\sigma_{LO}$, is plotted in Fig.~\ref{Yuk} vs.~the
Higgs-boson mass, $M_H$, at both the Tevatron and the LHC.  For values
of $M_H$ between 60 GeV and 1 TeV, the absolute value of the Yukawa
correction is never more than one percent of the leading-order cross
section.  Thus the Yukawa correction is negligible for this process,
as has also been found to be the case for $t\bar t$ production
\cite{STANGE,BDHMSW}. Since $W$-gluon fusion also involves the $t \bar
b$ weak charged current, our calculation suggests that the Yukawa
correction to that process is also negligible.  As previously
mentioned, the ordinary weak correction is expected to be comparable
to the Yukawa correction, so it too should be negligible.\footnote{The
complete calculation of the ordinary weak correction would require a
set of parton distribution functions which are extracted with weak
corrections included.  Such a set is not available at this time.} The
Yukawa correction could potentially be much larger in models with
enhanced couplings of Higgs bosons to top or bottom quarks
\cite{STANGE,KAO}.

\section{Conclusions}

\indent\indent The cross section for $q\bar q \to t\bar b, \bar t b$
at both the Tevatron and the LHC is given in Table \ref{table}.  The
leading-order cross section, next-to-leading-order cross section
including only the initial-state QCD correction, and the full
next-to-leading order cross section are given. The factorization and
renormalization scales are both set equal to $\sqrt{q^2}$, the mass of
the virtual $W$ boson.  We give results for three different sets of
next-to-leading-order parton distribution functions: CTEQ3M
\cite{CTEQ}, MRS(A$^{'}$), and MRS(G) \cite{MRSA}.\footnote{The
leading-order CTEQ cross section is calculated with the CTEQ3L
leading-order parton distribution functions.}  The QCD correction to
the cross section is significant: about $+54\%$ at the Tevatron and
$+50\%$ at the LHC, with the leading-order cross section evaluated
with leading-order parton distribution functions, and the
next-to-leading-order cross section evaluated with
next-to-leading-order parton distribution functions.\footnote{If
next-to-leading-order parton distribution functions are used at both
leading and next-to-leading order, the correction is about $+45\%$ at
the Tevatron and $+32\%$ at the LHC.}  The size of the ${\cal
O}(\alpha_s)$ correction improves the outlook for observation of this
process in Run II at the Tevatron.

\begin{table}
\begin{center}
\begin{tabular}{|c|c|c|c|c|} \hline
\multicolumn{2}{|c|}{$m_t = 175\;{\rm GeV}, \mu_R = \mu_F =
\sqrt{q^2}$} & \makebox[8em]{CTEQ3L,3M} & \makebox[8em]{MRS(A$'$)} &
\makebox[8em]{MRS(G)} \\ \hline & $\sigma_{LO}$ & .578 & .601 & .602
\\ \cline{2-5} Tevatron & $\sigma_{NLO(IS)}$ & .789 & .766 & .758 \\
\cline{2-5} $\sqrt{S}$ = 2 TeV & $\sigma_{NLO}$ & .894 & .868 & .860
\\ \cline{2-5} & \multicolumn{4}{|c|}{\bf\boldmath $\sigma_{\bf NLO}$
(avg) = .88 $\pm$ .05 pb} \\ \hline & $\sigma_{LO}$ & 6.76 & 7.83 &
7.81 \\ \cline {2-5} LHC & $\sigma_{NLO(IS)}$ & 9.02 & 9.01 & 9.03 \\
\cline{2-5} $\sqrt{S}$ = 14 TeV & $\sigma_{NLO}$ & 10.19 & 10.17 &
10.21 \\ \cline{2-5} & \multicolumn{4}{|c|}{\bf\boldmath $\sigma_{\bf
NLO}$ (avg) = 10.2 $\pm$ 0.6 pb} \\ \hline
\end{tabular}
\caption{\footnotesize Leading-order (LO) and next-to-leading-order
(NLO) cross sections (pb) for $q \bar q \to t \bar b,\bar t b$ at the
Tevatron and the LHC for three different sets of parton distribution
functions (PDFs).  The NLO cross section including only the initial
state (IS) correction is also given.  The CTEQ LO cross section is
computed with the CTEQ3L LO PDFs; all other cross sections are
computed with NLO PDFs.  The final NLO cross section is the average of
the CTEQ3M and MRS(A$'$) cross sections, with an uncertainty of $\pm
6\%$, as discussed in the text.}
\label{table}
\end{center}
\end{table}

As shown in Fig.~\ref{muf}, varying the factorization scale between
one half and twice $\sqrt {q^2}$ changes the cross section by only
$\pm 2\%$.  Varying the renormalization scale over this same range
yields a similar change in the cross section, as shown in
Fig.~\ref{mur}.  Using these results to estimate the contribution from
higher-order QCD corrections, we conclude that the uncertainty in the
cross section is at the level of $\pm 4\%$.  This conclusion is
supported by the known next-to-next-to-leading-order correction to the
Drell-Yan process, which is about $2\%$ (in the modified minimal
subtraction ($\overline{\rm MS}$) scheme) \cite{DY2}.

It is difficult to reliably ascertain the uncertainty in the cross
section from the parton distribution functions at this time.  The
small difference in the next-to-leading-order cross sections using
MRS(A$^{'}$) and MRS(G) supports the contention that the calculation
is insensitive to the gluon distribution function. The difference
between the cross section using CTEQ3M and MRS(A$^{'}$) suggests that
the uncertainty in the cross section from the parton distribution
functions is on the order of $\pm 2\%$.  However, since each set of
parton distribution functions represents the best fit to some set of
data, the uncertainty is certainly larger than this.  Therefore, we
assign an uncertainty of $\pm 4\%$ from the parton distribution
functions.

For our final estimate, we average the next-to-leading-order cross
sections using the CTEQ3M and MRS(A$^{'}$) parton distribution
functions.  We assign an uncertainty of $\pm 6\%$, which reflects the
uncertainties above, added in quadrature.  We quote as our final
result for $q\bar q\to t\bar b, \bar t b$ a cross section of $.88 \pm
.05$ pb at the Tevatron,\footnote{For $\sqrt{S} = 1.8$ TeV, the cross
section at the Tevatron is $.73 \pm .04$ pb} and $10.2 \pm 0.6$ pb at
the LHC.

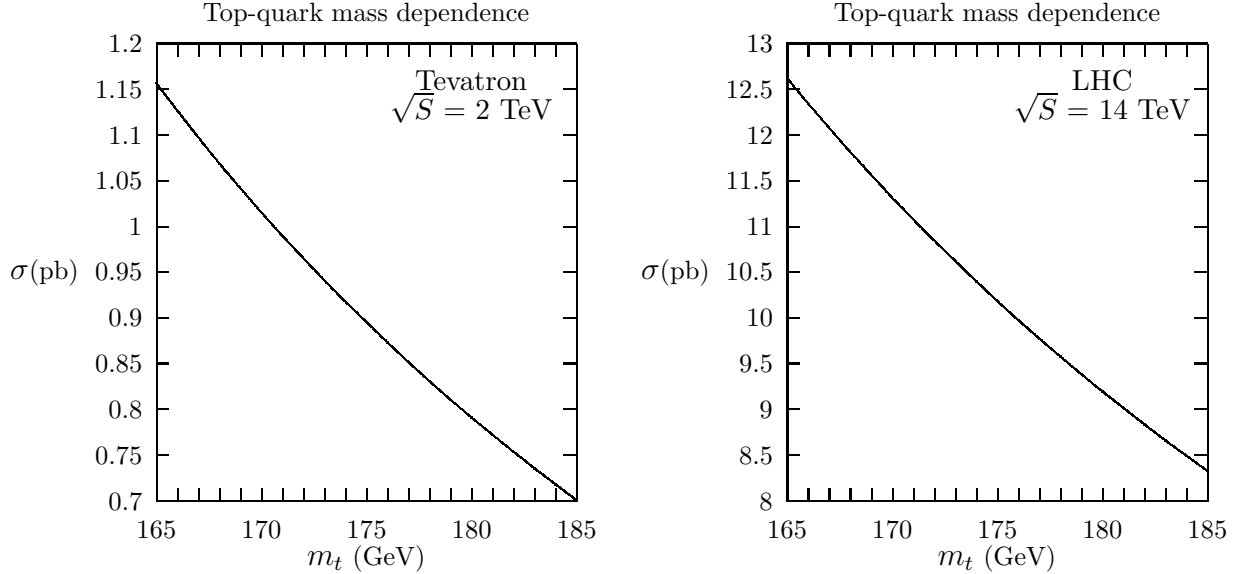
\begin{figure}
\setlength{\unitlength}{0.240900pt}
\ifx\plotpoint\undefined\newsavebox{\plotpoint}\fi
\begin{picture}(944,900)(0,0)
\font\gnuplot=cmr10 at 10pt
\gnuplot
\sbox{\plotpoint}{\rule[-0.200pt]{0.400pt}{0.400pt}}%
\put(220.0,113.0){\rule[-0.200pt]{4.818pt}{0.400pt}}
\put(198,113){\makebox(0,0)[r]{0.7}}
\put(860.0,113.0){\rule[-0.200pt]{4.818pt}{0.400pt}}
\put(220.0,185.0){\rule[-0.200pt]{4.818pt}{0.400pt}}
\put(198,185){\makebox(0,0)[r]{0.75}}
\put(860.0,185.0){\rule[-0.200pt]{4.818pt}{0.400pt}}
\put(220.0,257.0){\rule[-0.200pt]{4.818pt}{0.400pt}}
\put(198,257){\makebox(0,0)[r]{0.8}}
\put(860.0,257.0){\rule[-0.200pt]{4.818pt}{0.400pt}}
\put(220.0,329.0){\rule[-0.200pt]{4.818pt}{0.400pt}}
\put(198,329){\makebox(0,0)[r]{0.85}}
\put(860.0,329.0){\rule[-0.200pt]{4.818pt}{0.400pt}}
\put(220.0,401.0){\rule[-0.200pt]{4.818pt}{0.400pt}}
\put(198,401){\makebox(0,0)[r]{0.9}}
\put(860.0,401.0){\rule[-0.200pt]{4.818pt}{0.400pt}}
\put(220.0,473.0){\rule[-0.200pt]{4.818pt}{0.400pt}}
\put(198,473){\makebox(0,0)[r]{0.95}}
\put(860.0,473.0){\rule[-0.200pt]{4.818pt}{0.400pt}}
\put(220.0,544.0){\rule[-0.200pt]{4.818pt}{0.400pt}}
\put(198,544){\makebox(0,0)[r]{1}}
\put(860.0,544.0){\rule[-0.200pt]{4.818pt}{0.400pt}}
\put(220.0,616.0){\rule[-0.200pt]{4.818pt}{0.400pt}}
\put(198,616){\makebox(0,0)[r]{1.05}}
\put(860.0,616.0){\rule[-0.200pt]{4.818pt}{0.400pt}}
\put(220.0,688.0){\rule[-0.200pt]{4.818pt}{0.400pt}}
\put(198,688){\makebox(0,0)[r]{1.1}}
\put(860.0,688.0){\rule[-0.200pt]{4.818pt}{0.400pt}}
\put(220.0,760.0){\rule[-0.200pt]{4.818pt}{0.400pt}}
\put(198,760){\makebox(0,0)[r]{1.15}}
\put(860.0,760.0){\rule[-0.200pt]{4.818pt}{0.400pt}}
\put(220.0,832.0){\rule[-0.200pt]{4.818pt}{0.400pt}}
\put(198,832){\makebox(0,0)[r]{1.2}}
\put(860.0,832.0){\rule[-0.200pt]{4.818pt}{0.400pt}}
\put(220.0,113.0){\rule[-0.200pt]{0.400pt}{4.818pt}}
\put(220,68){\makebox(0,0){165}}
\put(220.0,812.0){\rule[-0.200pt]{0.400pt}{4.818pt}}
\put(253.0,113.0){\rule[-0.200pt]{0.400pt}{4.818pt}}
\put(253.0,812.0){\rule[-0.200pt]{0.400pt}{4.818pt}}
\put(286.0,113.0){\rule[-0.200pt]{0.400pt}{4.818pt}}
\put(286.0,812.0){\rule[-0.200pt]{0.400pt}{4.818pt}}
\put(319.0,113.0){\rule[-0.200pt]{0.400pt}{4.818pt}}
\put(319.0,812.0){\rule[-0.200pt]{0.400pt}{4.818pt}}
\put(352.0,113.0){\rule[-0.200pt]{0.400pt}{4.818pt}}
\put(352.0,812.0){\rule[-0.200pt]{0.400pt}{4.818pt}}
\put(385.0,113.0){\rule[-0.200pt]{0.400pt}{4.818pt}}
\put(385,68){\makebox(0,0){170}}
\put(385.0,812.0){\rule[-0.200pt]{0.400pt}{4.818pt}}
\put(418.0,113.0){\rule[-0.200pt]{0.400pt}{4.818pt}}
\put(418.0,812.0){\rule[-0.200pt]{0.400pt}{4.818pt}}
\put(451.0,113.0){\rule[-0.200pt]{0.400pt}{4.818pt}}
\put(451.0,812.0){\rule[-0.200pt]{0.400pt}{4.818pt}}
\put(484.0,113.0){\rule[-0.200pt]{0.400pt}{4.818pt}}
\put(484.0,812.0){\rule[-0.200pt]{0.400pt}{4.818pt}}
\put(517.0,113.0){\rule[-0.200pt]{0.400pt}{4.818pt}}
\put(517.0,812.0){\rule[-0.200pt]{0.400pt}{4.818pt}}
\put(550.0,113.0){\rule[-0.200pt]{0.400pt}{4.818pt}}
\put(550,68){\makebox(0,0){175}}
\put(550.0,812.0){\rule[-0.200pt]{0.400pt}{4.818pt}}
\put(583.0,113.0){\rule[-0.200pt]{0.400pt}{4.818pt}}
\put(583.0,812.0){\rule[-0.200pt]{0.400pt}{4.818pt}}
\put(616.0,113.0){\rule[-0.200pt]{0.400pt}{4.818pt}}
\put(616.0,812.0){\rule[-0.200pt]{0.400pt}{4.818pt}}
\put(649.0,113.0){\rule[-0.200pt]{0.400pt}{4.818pt}}
\put(649.0,812.0){\rule[-0.200pt]{0.400pt}{4.818pt}}
\put(682.0,113.0){\rule[-0.200pt]{0.400pt}{4.818pt}}
\put(682.0,812.0){\rule[-0.200pt]{0.400pt}{4.818pt}}
\put(715.0,113.0){\rule[-0.200pt]{0.400pt}{4.818pt}}
\put(715,68){\makebox(0,0){180}}
\put(715.0,812.0){\rule[-0.200pt]{0.400pt}{4.818pt}}
\put(748.0,113.0){\rule[-0.200pt]{0.400pt}{4.818pt}}
\put(748.0,812.0){\rule[-0.200pt]{0.400pt}{4.818pt}}
\put(781.0,113.0){\rule[-0.200pt]{0.400pt}{4.818pt}}
\put(781.0,812.0){\rule[-0.200pt]{0.400pt}{4.818pt}}
\put(814.0,113.0){\rule[-0.200pt]{0.400pt}{4.818pt}}
\put(814.0,812.0){\rule[-0.200pt]{0.400pt}{4.818pt}}
\put(847.0,113.0){\rule[-0.200pt]{0.400pt}{4.818pt}}
\put(847.0,812.0){\rule[-0.200pt]{0.400pt}{4.818pt}}
\put(880.0,113.0){\rule[-0.200pt]{0.400pt}{4.818pt}}
\put(880,68){\makebox(0,0){185}}
\put(880.0,812.0){\rule[-0.200pt]{0.400pt}{4.818pt}}
\put(220.0,113.0){\rule[-0.200pt]{158.994pt}{0.400pt}}
\put(880.0,113.0){\rule[-0.200pt]{0.400pt}{173.207pt}}
\put(220.0,832.0){\rule[-0.200pt]{158.994pt}{0.400pt}}
\put(45,472){\makebox(0,0){$\sigma$(pb)}}
\put(550,23){\makebox(0,0){$m_t$ (GeV)}}
\put(550,877){\makebox(0,0){Top-quark mass dependence}}
\put(715,774){\makebox(0,0){\small Tevatron}}
\put(715,731){\makebox(0,0){\small $\sqrt{S}$ = 2 TeV}}
\put(220.0,113.0){\rule[-0.200pt]{0.400pt}{173.207pt}}
\put(220,770){\usebox{\plotpoint}}
\multiput(220.58,767.42)(0.497,-0.652){63}{\rule{0.120pt}{0.621pt}}
\multiput(219.17,768.71)(33.000,-41.711){2}{\rule{0.400pt}{0.311pt}}
\multiput(253.58,724.47)(0.497,-0.637){63}{\rule{0.120pt}{0.609pt}}
\multiput(252.17,725.74)(33.000,-40.736){2}{\rule{0.400pt}{0.305pt}}
\multiput(286.58,682.52)(0.497,-0.621){63}{\rule{0.120pt}{0.597pt}}
\multiput(285.17,683.76)(33.000,-39.761){2}{\rule{0.400pt}{0.298pt}}
\multiput(319.58,641.62)(0.497,-0.591){63}{\rule{0.120pt}{0.573pt}}
\multiput(318.17,642.81)(33.000,-37.811){2}{\rule{0.400pt}{0.286pt}}
\multiput(352.58,602.67)(0.497,-0.575){63}{\rule{0.120pt}{0.561pt}}
\multiput(351.17,603.84)(33.000,-36.836){2}{\rule{0.400pt}{0.280pt}}
\multiput(385.58,564.72)(0.497,-0.560){63}{\rule{0.120pt}{0.548pt}}
\multiput(384.17,565.86)(33.000,-35.862){2}{\rule{0.400pt}{0.274pt}}
\multiput(418.58,527.82)(0.497,-0.529){63}{\rule{0.120pt}{0.524pt}}
\multiput(417.17,528.91)(33.000,-33.912){2}{\rule{0.400pt}{0.262pt}}
\multiput(451.58,492.82)(0.497,-0.529){63}{\rule{0.120pt}{0.524pt}}
\multiput(450.17,493.91)(33.000,-33.912){2}{\rule{0.400pt}{0.262pt}}
\multiput(484.58,457.87)(0.497,-0.514){63}{\rule{0.120pt}{0.512pt}}
\multiput(483.17,458.94)(33.000,-32.937){2}{\rule{0.400pt}{0.256pt}}
\multiput(517.00,424.92)(0.515,-0.497){61}{\rule{0.512pt}{0.120pt}}
\multiput(517.00,425.17)(31.936,-32.000){2}{\rule{0.256pt}{0.400pt}}
\multiput(550.00,392.92)(0.515,-0.497){61}{\rule{0.512pt}{0.120pt}}
\multiput(550.00,393.17)(31.936,-32.000){2}{\rule{0.256pt}{0.400pt}}
\multiput(583.00,360.92)(0.531,-0.497){59}{\rule{0.526pt}{0.120pt}}
\multiput(583.00,361.17)(31.909,-31.000){2}{\rule{0.263pt}{0.400pt}}
\multiput(616.00,329.92)(0.549,-0.497){57}{\rule{0.540pt}{0.120pt}}
\multiput(616.00,330.17)(31.879,-30.000){2}{\rule{0.270pt}{0.400pt}}
\multiput(649.00,299.92)(0.568,-0.497){55}{\rule{0.555pt}{0.120pt}}
\multiput(649.00,300.17)(31.848,-29.000){2}{\rule{0.278pt}{0.400pt}}
\multiput(682.00,270.92)(0.589,-0.497){53}{\rule{0.571pt}{0.120pt}}
\multiput(682.00,271.17)(31.814,-28.000){2}{\rule{0.286pt}{0.400pt}}
\multiput(715.00,242.92)(0.611,-0.497){51}{\rule{0.589pt}{0.120pt}}
\multiput(715.00,243.17)(31.778,-27.000){2}{\rule{0.294pt}{0.400pt}}
\multiput(748.00,215.92)(0.611,-0.497){51}{\rule{0.589pt}{0.120pt}}
\multiput(748.00,216.17)(31.778,-27.000){2}{\rule{0.294pt}{0.400pt}}
\multiput(781.00,188.92)(0.635,-0.497){49}{\rule{0.608pt}{0.120pt}}
\multiput(781.00,189.17)(31.739,-26.000){2}{\rule{0.304pt}{0.400pt}}
\multiput(814.00,162.92)(0.661,-0.497){47}{\rule{0.628pt}{0.120pt}}
\multiput(814.00,163.17)(31.697,-25.000){2}{\rule{0.314pt}{0.400pt}}
\multiput(847.00,137.92)(0.661,-0.497){47}{\rule{0.628pt}{0.120pt}}
\multiput(847.00,138.17)(31.697,-25.000){2}{\rule{0.314pt}{0.400pt}}
\put(880,114){\usebox{\plotpoint}}
\end{picture}
\setlength{\unitlength}{0.240900pt}
\ifx\plotpoint\undefined\newsavebox{\plotpoint}\fi
\begin{picture}(944,900)(0,0)
\font\gnuplot=cmr10 at 10pt
\gnuplot
\sbox{\plotpoint}{\rule[-0.200pt]{0.400pt}{0.400pt}}%
\put(220.0,113.0){\rule[-0.200pt]{4.818pt}{0.400pt}}
\put(198,113){\makebox(0,0)[r]{8}}
\put(860.0,113.0){\rule[-0.200pt]{4.818pt}{0.400pt}}
\put(220.0,185.0){\rule[-0.200pt]{4.818pt}{0.400pt}}
\put(198,185){\makebox(0,0)[r]{8.5}}
\put(860.0,185.0){\rule[-0.200pt]{4.818pt}{0.400pt}}
\put(220.0,257.0){\rule[-0.200pt]{4.818pt}{0.400pt}}
\put(198,257){\makebox(0,0)[r]{9}}
\put(860.0,257.0){\rule[-0.200pt]{4.818pt}{0.400pt}}
\put(220.0,329.0){\rule[-0.200pt]{4.818pt}{0.400pt}}
\put(198,329){\makebox(0,0)[r]{9.5}}
\put(860.0,329.0){\rule[-0.200pt]{4.818pt}{0.400pt}}
\put(220.0,401.0){\rule[-0.200pt]{4.818pt}{0.400pt}}
\put(198,401){\makebox(0,0)[r]{10}}
\put(860.0,401.0){\rule[-0.200pt]{4.818pt}{0.400pt}}
\put(220.0,473.0){\rule[-0.200pt]{4.818pt}{0.400pt}}
\put(198,473){\makebox(0,0)[r]{10.5}}
\put(860.0,473.0){\rule[-0.200pt]{4.818pt}{0.400pt}}
\put(220.0,544.0){\rule[-0.200pt]{4.818pt}{0.400pt}}
\put(198,544){\makebox(0,0)[r]{11}}
\put(860.0,544.0){\rule[-0.200pt]{4.818pt}{0.400pt}}
\put(220.0,616.0){\rule[-0.200pt]{4.818pt}{0.400pt}}
\put(198,616){\makebox(0,0)[r]{11.5}}
\put(860.0,616.0){\rule[-0.200pt]{4.818pt}{0.400pt}}
\put(220.0,688.0){\rule[-0.200pt]{4.818pt}{0.400pt}}
\put(198,688){\makebox(0,0)[r]{12}}
\put(860.0,688.0){\rule[-0.200pt]{4.818pt}{0.400pt}}
\put(220.0,760.0){\rule[-0.200pt]{4.818pt}{0.400pt}}
\put(198,760){\makebox(0,0)[r]{12.5}}
\put(860.0,760.0){\rule[-0.200pt]{4.818pt}{0.400pt}}
\put(220.0,832.0){\rule[-0.200pt]{4.818pt}{0.400pt}}
\put(198,832){\makebox(0,0)[r]{13}}
\put(860.0,832.0){\rule[-0.200pt]{4.818pt}{0.400pt}}
\put(220.0,113.0){\rule[-0.200pt]{0.400pt}{4.818pt}}
\put(220,68){\makebox(0,0){165}}
\put(220.0,812.0){\rule[-0.200pt]{0.400pt}{4.818pt}}
\put(253.0,113.0){\rule[-0.200pt]{0.400pt}{4.818pt}}
\put(253.0,812.0){\rule[-0.200pt]{0.400pt}{4.818pt}}
\put(286.0,113.0){\rule[-0.200pt]{0.400pt}{4.818pt}}
\put(286.0,812.0){\rule[-0.200pt]{0.400pt}{4.818pt}}
\put(319.0,113.0){\rule[-0.200pt]{0.400pt}{4.818pt}}
\put(319.0,812.0){\rule[-0.200pt]{0.400pt}{4.818pt}}
\put(352.0,113.0){\rule[-0.200pt]{0.400pt}{4.818pt}}
\put(352.0,812.0){\rule[-0.200pt]{0.400pt}{4.818pt}}
\put(385.0,113.0){\rule[-0.200pt]{0.400pt}{4.818pt}}
\put(385,68){\makebox(0,0){170}}
\put(385.0,812.0){\rule[-0.200pt]{0.400pt}{4.818pt}}
\put(418.0,113.0){\rule[-0.200pt]{0.400pt}{4.818pt}}
\put(418.0,812.0){\rule[-0.200pt]{0.400pt}{4.818pt}}
\put(451.0,113.0){\rule[-0.200pt]{0.400pt}{4.818pt}}
\put(451.0,812.0){\rule[-0.200pt]{0.400pt}{4.818pt}}
\put(484.0,113.0){\rule[-0.200pt]{0.400pt}{4.818pt}}
\put(484.0,812.0){\rule[-0.200pt]{0.400pt}{4.818pt}}
\put(517.0,113.0){\rule[-0.200pt]{0.400pt}{4.818pt}}
\put(517.0,812.0){\rule[-0.200pt]{0.400pt}{4.818pt}}
\put(550.0,113.0){\rule[-0.200pt]{0.400pt}{4.818pt}}
\put(550,68){\makebox(0,0){175}}
\put(550.0,812.0){\rule[-0.200pt]{0.400pt}{4.818pt}}
\put(583.0,113.0){\rule[-0.200pt]{0.400pt}{4.818pt}}
\put(583.0,812.0){\rule[-0.200pt]{0.400pt}{4.818pt}}
\put(616.0,113.0){\rule[-0.200pt]{0.400pt}{4.818pt}}
\put(616.0,812.0){\rule[-0.200pt]{0.400pt}{4.818pt}}
\put(649.0,113.0){\rule[-0.200pt]{0.400pt}{4.818pt}}
\put(649.0,812.0){\rule[-0.200pt]{0.400pt}{4.818pt}}
\put(682.0,113.0){\rule[-0.200pt]{0.400pt}{4.818pt}}
\put(682.0,812.0){\rule[-0.200pt]{0.400pt}{4.818pt}}
\put(715.0,113.0){\rule[-0.200pt]{0.400pt}{4.818pt}}
\put(715,68){\makebox(0,0){180}}
\put(715.0,812.0){\rule[-0.200pt]{0.400pt}{4.818pt}}
\put(748.0,113.0){\rule[-0.200pt]{0.400pt}{4.818pt}}
\put(748.0,812.0){\rule[-0.200pt]{0.400pt}{4.818pt}}
\put(781.0,113.0){\rule[-0.200pt]{0.400pt}{4.818pt}}
\put(781.0,812.0){\rule[-0.200pt]{0.400pt}{4.818pt}}
\put(814.0,113.0){\rule[-0.200pt]{0.400pt}{4.818pt}}
\put(814.0,812.0){\rule[-0.200pt]{0.400pt}{4.818pt}}
\put(847.0,113.0){\rule[-0.200pt]{0.400pt}{4.818pt}}
\put(847.0,812.0){\rule[-0.200pt]{0.400pt}{4.818pt}}
\put(880.0,113.0){\rule[-0.200pt]{0.400pt}{4.818pt}}
\put(880,68){\makebox(0,0){185}}
\put(880.0,812.0){\rule[-0.200pt]{0.400pt}{4.818pt}}
\put(220.0,113.0){\rule[-0.200pt]{158.994pt}{0.400pt}}
\put(880.0,113.0){\rule[-0.200pt]{0.400pt}{173.207pt}}
\put(220.0,832.0){\rule[-0.200pt]{158.994pt}{0.400pt}}
\put(45,472){\makebox(0,0){$\sigma$(pb)}}
\put(550,23){\makebox(0,0){$m_t$ (GeV)}}
\put(550,877){\makebox(0,0){Top-quark mass dependence}}
\put(715,774){\makebox(0,0){\small LHC}}
\put(715,731){\makebox(0,0){\small $\sqrt{S}$ = 14 TeV}}
\put(220.0,113.0){\rule[-0.200pt]{0.400pt}{173.207pt}}
\put(220,778){\usebox{\plotpoint}}
\multiput(220.58,775.57)(0.497,-0.606){63}{\rule{0.120pt}{0.585pt}}
\multiput(219.17,776.79)(33.000,-38.786){2}{\rule{0.400pt}{0.292pt}}
\multiput(253.58,735.67)(0.497,-0.575){63}{\rule{0.120pt}{0.561pt}}
\multiput(252.17,736.84)(33.000,-36.836){2}{\rule{0.400pt}{0.280pt}}
\multiput(286.58,697.67)(0.497,-0.575){63}{\rule{0.120pt}{0.561pt}}
\multiput(285.17,698.84)(33.000,-36.836){2}{\rule{0.400pt}{0.280pt}}
\multiput(319.58,659.77)(0.497,-0.545){63}{\rule{0.120pt}{0.536pt}}
\multiput(318.17,660.89)(33.000,-34.887){2}{\rule{0.400pt}{0.268pt}}
\multiput(352.58,623.77)(0.497,-0.545){63}{\rule{0.120pt}{0.536pt}}
\multiput(351.17,624.89)(33.000,-34.887){2}{\rule{0.400pt}{0.268pt}}
\multiput(385.58,587.87)(0.497,-0.514){63}{\rule{0.120pt}{0.512pt}}
\multiput(384.17,588.94)(33.000,-32.937){2}{\rule{0.400pt}{0.256pt}}
\multiput(418.58,553.87)(0.497,-0.514){63}{\rule{0.120pt}{0.512pt}}
\multiput(417.17,554.94)(33.000,-32.937){2}{\rule{0.400pt}{0.256pt}}
\multiput(451.00,520.92)(0.515,-0.497){61}{\rule{0.512pt}{0.120pt}}
\multiput(451.00,521.17)(31.936,-32.000){2}{\rule{0.256pt}{0.400pt}}
\multiput(484.00,488.92)(0.515,-0.497){61}{\rule{0.512pt}{0.120pt}}
\multiput(484.00,489.17)(31.936,-32.000){2}{\rule{0.256pt}{0.400pt}}
\multiput(517.00,456.92)(0.531,-0.497){59}{\rule{0.526pt}{0.120pt}}
\multiput(517.00,457.17)(31.909,-31.000){2}{\rule{0.263pt}{0.400pt}}
\multiput(550.00,425.92)(0.549,-0.497){57}{\rule{0.540pt}{0.120pt}}
\multiput(550.00,426.17)(31.879,-30.000){2}{\rule{0.270pt}{0.400pt}}
\multiput(583.00,395.92)(0.568,-0.497){55}{\rule{0.555pt}{0.120pt}}
\multiput(583.00,396.17)(31.848,-29.000){2}{\rule{0.278pt}{0.400pt}}
\multiput(616.00,366.92)(0.589,-0.497){53}{\rule{0.571pt}{0.120pt}}
\multiput(616.00,367.17)(31.814,-28.000){2}{\rule{0.286pt}{0.400pt}}
\multiput(649.00,338.92)(0.589,-0.497){53}{\rule{0.571pt}{0.120pt}}
\multiput(649.00,339.17)(31.814,-28.000){2}{\rule{0.286pt}{0.400pt}}
\multiput(682.00,310.92)(0.611,-0.497){51}{\rule{0.589pt}{0.120pt}}
\multiput(682.00,311.17)(31.778,-27.000){2}{\rule{0.294pt}{0.400pt}}
\multiput(715.00,283.92)(0.635,-0.497){49}{\rule{0.608pt}{0.120pt}}
\multiput(715.00,284.17)(31.739,-26.000){2}{\rule{0.304pt}{0.400pt}}
\multiput(748.00,257.92)(0.635,-0.497){49}{\rule{0.608pt}{0.120pt}}
\multiput(748.00,258.17)(31.739,-26.000){2}{\rule{0.304pt}{0.400pt}}
\multiput(781.00,231.92)(0.661,-0.497){47}{\rule{0.628pt}{0.120pt}}
\multiput(781.00,232.17)(31.697,-25.000){2}{\rule{0.314pt}{0.400pt}}
\multiput(814.00,206.92)(0.689,-0.496){45}{\rule{0.650pt}{0.120pt}}
\multiput(814.00,207.17)(31.651,-24.000){2}{\rule{0.325pt}{0.400pt}}
\multiput(847.00,182.92)(0.689,-0.496){45}{\rule{0.650pt}{0.120pt}}
\multiput(847.00,183.17)(31.651,-24.000){2}{\rule{0.325pt}{0.400pt}}
\put(880,160){\usebox{\plotpoint}}
\end{picture}
\caption{\footnotesize Next-to-leading order cross section for $q\bar
q\to t\bar b, \bar t b$ as a function of the top-quark mass.}
\label{mt}
\end{figure}

An additional source of uncertainty is the top-quark mass.  A plot of
the next-to-leading order cross section for $q\bar q\to t\bar b, \bar
t b$ as a function of the top-quark mass is show in Fig.~\ref{mt}.  It
is anticipated that this uncertainty will be $\pm 6$ GeV when the data
from Run I at the Tevatron are fully analyzed.  This yields an
uncertainty of $\pm 15\%$ in the cross section at the Tevatron.  The
uncertainty in the mass is expected to decrease to $\pm 4$ GeV in Run
II \cite{tev2000}, corresponding to an uncertainty in the cross
section of $\pm 10\%$.  A high-luminosity Tevatron might be capable of
reducing the uncertainty in the mass to $\pm 2$ GeV \cite{tev2000},
which would yield an uncertainty in the cross section of $\pm 5\%$.
The uncertainty in the cross section at the LHC is comparable.

Much can be done to reduce the uncertainty in the calculation.  The
next-to-next-to-leading-order correction to the Drell-Yan process is
already known \cite{DY2}.  The full next-to-next-to-leading-order QCD
correction to $q \bar q \to t \bar b$ can and should be completed in
the near future.  This should reduce the uncertainty in the cross
section from yet higher orders to below the $1\%$ level.  A reliable
estimate of the uncertainty in the parton distribution functions
requires a set with built-in uncertainties, which we hope will be
available in the near future.

It seems likely that by the time the process $q\bar q\to t\bar b$ is
observed in Run II at the Tevatron, the theoretical uncertainty in the
cross section will be slightly larger than $\pm 10\%$, due mostly to
the uncertainty in the mass.  This is adequate in comparison with the
anticipated experimental errors.  The statistical error on the
measured cross section in Run II will be about $\pm 20\%$ \cite{SW}.
This corresponds to a measurement of $|V_{tb}|$ with an accuracy of
$\pm 10\%$ (assuming $|V_{tb}| \approx 1$).  A high-luminosity
Tevatron, which could potentially deliver 30 fb$^{-1}$ over several
years, would allow a measurement of the cross section with a
statistical uncertainty of about $6\%$, with a comparable theoretical
uncertainty.  Combining the statistical and theoretical uncertainties
in quadrature, this corresponds to a measurement of $|V_{tb}|$ with an
accuracy of about $\pm 4\%$.

\begin{figure}
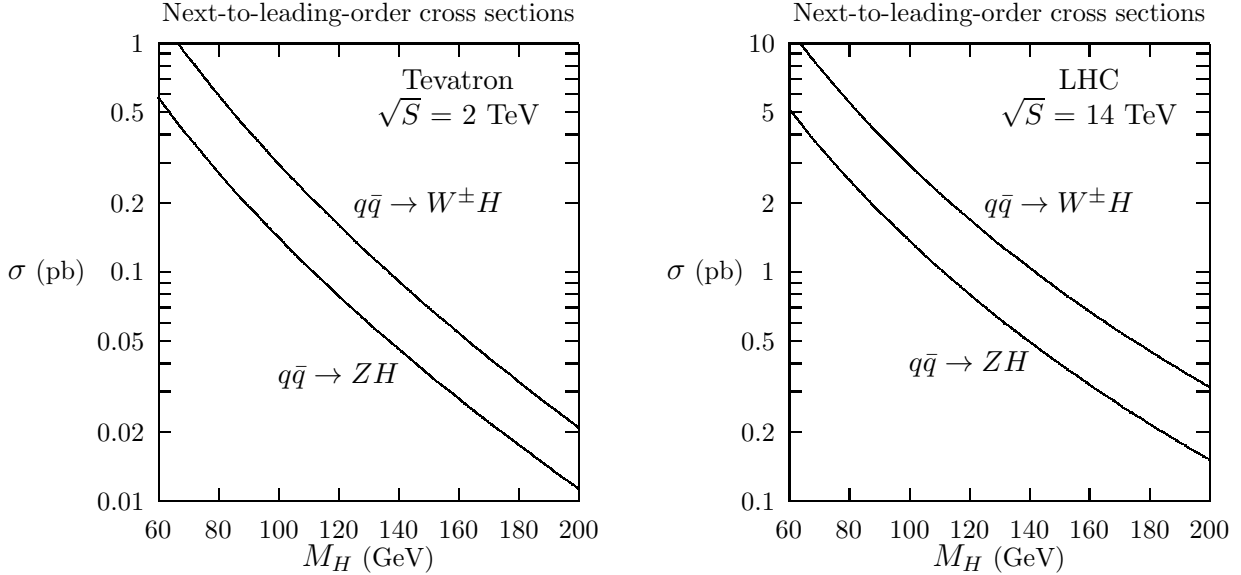

\input{wh_tev.tex}
\input{wh_lhc.tex}
\caption{\footnotesize Cross section for $q \bar q \to W^{\pm}H$ and
$q \bar q \to ZH$ at the Tevatron and the LHC, calculated at
next-to-leading order using CTEQ3M parton distribution functions.}
\label{wh}
\end{figure}

The process $q\bar q \to t\bar b$ is also important as a background to
the process $q\bar q\to WH$ with $H\to b\bar b$ at the Tevatron.  We
show in Fig.~\ref{wh} the next-to-leading-order cross sections for
$q\bar q \to W^{\pm}H$, as well as $q\bar q \to ZH$, at both the
Tevatron and the LHC \cite{HW}.  The significant increase in the
$q\bar q\to t\bar b$ cross section at next-to-leading order could have
a negative impact on the ability to find an intermediate-mass Higgs
boson at the Tevatron.

\section*{Acknowledgements}

\indent\indent We are grateful for conversations with S.~Keller,
B.~Kniehl, S.~Kuhlmann, and T.~Stelzer. This work was supported in
part by Department of Energy grant DE-FG02-91ER40677.  We gratefully
acknowledge the support of a GAANN fellowship, under grant number
DE-P200A40532 from the U.~S.~Department of Education for MS.

\section*{Appendix}

\indent\indent Below are the form factors for the Yukawa correction to
the matrix element of the $t \bar b$ charged current.  These
corrections arise from loops of Higgs bosons and the unphysical scalar
$W$ and $Z$ bosons associated with the Higgs field in the $R_{\xi}$
gauges.  The masses-squared of the unphysical scalar bosons are $\xi
M_W^2,\xi M_Z^2$.  In the numerical calculations, we set $\xi = 0$
(Landau gauge). The integrals were reduced to the standard one-, two-,
and three-point scalar loop integrals and then evaluated with the aid
of the code FF \cite{FF}.  The notation is adopted from \cite{PV}; the
arguments of the functions give the internal masses-squared followed
by the external momenta squared.

\begin{eqnarray*}
F_1 & = & {\small \frac{1}{2}} [4C_{24}(\xi
M_W^2,M_H^2,m_t^2;q^2,m_t^2,0) + 4C_{24}(\xi M_W^2,\xi
M_Z^2,m_t^2;q^2,m_t^2,0) \\ & & \mbox{} + B_1(M_H^2,m_t^2;m_t^2) +
(M_H^2 - 4m_t^2) B_0^{'}(M_H^2,m_t^2;m_t^2) \\ & & \mbox{} + B_1(\xi
M_Z^2,m_t^2;m_t^2) + \xi M_Z^2 B_0^{'} (\xi M_Z^2,m_t^2;m_t^2) \\ & &
\mbox{} + B_1(\xi M_W^2,0;m_t^2) + (\xi M_W^2 - m_t^2) B_0^{'}(\xi
M_W^2,0;m_t^2) \\ & & \mbox{} + B_1(\xi M_W^2,m_t^2;0) + (\xi M_W^2 +
m_t^2) B_0^{'}(\xi M_W^2,m_t^2;0)] \\ &&\\ F_2 & = & m_t^2 [C_{23}(\xi
M_W^2,M_H^2,m_t^2;q^2,m_t^2,0) + C_{23}(\xi M_W^2,\xi
M_Z^2,m_t^2;q^2,m_t^2,0) \\ & & \mbox{} + C_{21}(\xi
M_W^2,M_H^2,m_t^2;q^2,m_t^2,0) + C_{21}(\xi M_W^2,\xi
M_Z^2,m_t^2;q^2,m_t^2,0) \\ & & \mbox{} + 2C_{11}(\xi
M_W^2,M_H^2,m_t^2;q^2,m_t^2,0)]
\end{eqnarray*}

Note: In the reduction of the three-point integrals, a misprint was
discovered in Ref.~\cite{PV}.  On p.~199 in Appendix E, in an
unnumbered equation near the bottom of the page, $C_{22}$ and $C_{23}$
were transposed.  The correct equation is \mbox{$(C_{23},C_{22}) =
X^{-1}(R_4,R_6)$}.

\end{document}